\documentclass[aps,showpacs,preprint,epsf,epsfig, nofootinbib]{revtex4-1}
\usepackage{ifxetex,ifluatex}
\if\ifxetex T\else\ifluatex T\else F\fi\fi T%
\usepackage{fontspec}
\else
\usepackage[T1]{fontenc}
\usepackage[utf8]{inputenc}
\usepackage{lmodern}
\fi
\usepackage{mathrsfs}
\usepackage{dcolumn}
\usepackage{bm}
\usepackage{hyperref}
\usepackage{epsfig,graphicx,color,appendix}
\usepackage{multirow}
\usepackage{tabularx}
\usepackage{capt-of}
\usepackage{caption}
\usepackage{multirow}
\usepackage[countmax]{subfloat}
\usepackage[utf8]{inputenc}
\usepackage{graphicx}
\renewcommand{\baselinestretch}{1.10}
\usepackage{threeparttable}
\usepackage{amssymb}
\usepackage{array}
\usepackage{amsmath}
\usepackage{mathtools}
\renewcommand{\thefootnote}{\fnsymbol{footnote}}
\usepackage[english]{babel}
\begin{document}
	\def\ben{\begin{eqnarray}}
	\def\ben{\end{eqnarray}}
	\def\non{\nonumber}
	\def\obt{\frac{1}{2}}
	\def\tbt{\frac{3}{2}}
	\def\obtp{\frac{1^{\prime}}{2}}
	\def\la{\langle}
	\def\ra{\rangle}
	\def\t{\times}
	\def\pp{{\prime\prime}}
	\def\nc{N_c^{\rm eff}}
	\def\vp{\varepsilon}
	\def\hep{\hat{\varepsilon}}
	\def\a{{\cal A}}
	\def\B{{\cal B}}
	\def\c{{\cal C}}
	\def\d{{\cal D}}
	\def\e{{\cal E}}
	\def\p{{\cal P}}
	\def\tt{{\cal T}}
	\def\up{\uparrow}
	\def\dw{\downarrow}
	\def\vma{{_{V-A}}}
	\def\vpa{{_{V+A}}}
	\def\smp{{_{S-P}}}
	\def\spp{{_{S+P}}}
	\def\J{{J/\psi}}
	\def\ov{\overline}
	\def\Lqcd{{\Lambda_{\rm QCD}}}
	\def\pr{{Phys. Rev.}~}
	\def\prl{{ Phys. Rev. Lett.}~}
	\def\pl{{ Phys. Lett.}~}
	\def\np{{ Nucl. Phys.}~}
	\def\zp{{ Z. Phys.}~}
	\def\plus{\texttt{+}}
	\def\minus{\texttt{-}}
	\renewcommand{\baselinestretch}{1.10}
	\long\def\symbolfootnote[#1]#2{\begingroup%
		\def\thefootnote{\fnsymbol{footnote}}\footnote[#1]{#2}\endgroup}
	\def\lsim{ {\ \lower-1.2pt\vbox{\hbox{\rlap{$<$}\lower5pt\vbox{\hbox{$\sim$}
			}}}\ } }
	\def\gsim{ {\ \lower-1.2pt\vbox{\hbox{\rlap{$>$}\lower5pt\vbox{\hbox{$\sim$}
			}}}\ } }
	\font\el=cmbx10 scaled \magstep2{\obeylines\hfill \today}
	
	\vskip 1.5 cm
	\centerline{\large\bf Radiative M1 transitions of heavy baryons:  }
	\centerline{\large\bf Effective Quark Mass Scheme}
	
	\small
	\vskip 1.0 cm
	
	\centerline{\bf Avijit Hazra\footnote[1]{\href{mailto:avijithr@srmist.edu.in}{avijithr@srmist.edu.in}}, Saheli Rakshit\footnote[2]{\href{mailto:sahelirakshit@gmail.com}{sahelirakshit@gmail.com}}, and Rohit Dhir\footnote[3]{Corresponding author: \href{mailto:dhir.rohit@gmail.com}{dhir.rohit@gmail.com}}}
	\medskip
	\centerline{\it Department of Physics and Nanotechnology,}
	\centerline{\it SRM Institute of Science and Technology, Kattankulathur 603203, India.}
	\bigskip
	\bigskip
	\begin{center}
		{\large \bf Abstract}
	\end{center}
  We calculate the magnetic moments of ground state $J^P=\frac{1}{2}^+$ and $J^P=\frac{3}{2}^+$ heavy flavor charm and bottom baryon states employing the concept of effective mass based on single gluon exchange interaction coupling to the spectator quarks in the non-relativistic quark model. We exploit the current experimental information in the heavy flavor sector to estimate the interaction contributions to get the effective masses of the quarks inside the baryons. We study the spin  $\frac{1}{2}^{'+} \rightarrow  \frac{1}{2}^+$, $\frac{3}{2}^+ \rightarrow \frac{1}{2}^+$, and $\frac{3}{2}^+ \rightarrow \frac{1}{2}^{'+}$ transition moments for these baryons. We make robust predictions of the radiative M1 decay widths of singly, doubly, and triply heavy flavored baryons. \\
	
	\medskip
	\medskip
	

	\vfill
	\section{Introduction}
	  In the past few years, the LHCb has been on a discovering spree and has reported a large number of heavy flavor baryons and resonances, which has gathered a formidable attention from physicists. In fact, these experimental observations have presented a number of scenarios to investigate heavy flavor baryon mass spectra and their structure. In recent years, the LHCb collaboration has reported the observation of five new narrow $\Omega_c^{0}$ states, doubly charm $\Xi_{cc}^{++}$, excited states of $\Sigma_b(6097)$, bottom-strange state $\Xi_b(6227)$, resonances $\Lambda_b(6146)$, and $\Lambda_b(6152)$  \cite{Aaij:2017nav, Aaij:2017ueg, Aaij:2018wzf, Aaij:2018yqz, Aaij:2018tnn, Aaij:2019amv, Sirunyan:2020gtz, Aaij:2020rkw}. The LHCb Collaboration has also discovered four narrow peaks for $\Omega_b^-$ excited states \cite{Aaij:2020cex}. Most recently, the CMS Collaboration has found a new state $\Xi_b(6100)$ \cite{Sirunyan:2021vxz}. Many spin-$\frac{1}{2}$ \textit{b}-baryons ($\Lambda_b^0$, $\Sigma_b^{\pm,0}$, $\Xi_b^{0,-}$, $\Xi_b^{'-}$, $\Omega_b^{0}$), and spin-$\frac{3}{2}$ baryons  ($\Lambda_{b}^{*0}$, $\Sigma_b^{*\pm,0}$, $\Xi_b^{*0,-}$) have also been discovered, and their masses have been measured precisely \cite{Aaij:2016jnn, Aaij:2017awb, Aaij:2019ezy, Zyla:2020zbs,Chen:2016spr}. Furthermore, the LHCb has already been attempting the searches of doubly heavy bottom charmed baryons $\Xi_{cb}$ and is committed to improve the signal sensitivity for these states \cite{Aaij:2020vid}. This tremendous experimental activity has motivated a number of theorists and phenomenologists to explore the structural properties of heavy flavor baryons. A complete understanding of the hadrons as a bound state of quarks and gluons, and confinement, their spectra and internal structure, the heavy-quark dynamics remain fundamental challenges of hadronic physics, and Quantum Chromodynamics (QCD) at large \cite{DeRujula:1975qlm,Brambilla:2014jmp}. Understanding electromagnetic properties and radiative transitions of heavy baryons are the keys to heavy flavor dynamics. In this process, observation of the heavy flavor spectroscopy, magnetic moments, and transition decay widths provide the testing hypothesis for various theoretical models on the hadron structure.
	  
	  In this work, we have focused on transition moments and radiative M1 decay widths of heavy flavor baryons. The studies of transition moments and consequently, M1 decay widths are important when phase space forbids strong decay channels. Experimentally, three of the radiative decay processes, $\Omega_{c}^{*}\rightarrow\Omega_{c}\gamma$, $\Xi_{c}^{'+}\rightarrow\Xi_{c}^{+}\gamma$, and $\Xi_c^{'0}\rightarrow\Xi_{c}^{0}\gamma$  have been observed by $\it{BABAR}$ and Belle Collaborations \cite{Aubert:2006je, Solovieva:2008fw, Jessop:1998wt, Aubert:2006rv, Yelton:2016fqw}. One can expect more experimental results in the near future on the radiative decay widths of charm and bottom baryons from the upgraded experimental facilities of BES-III and LHCb \cite{Ablikim:2019hff,Yuan:2019zfo,Aiola:2020yam,Fomin:2019wuw,Audurier:2021wqk}. On the other hand, the experimental progress on the measurement of baryon magnetic moments has been very slow. So far, magnetic moments of all the low lying $J^P=\frac{1}{2}^+$ octet baryons, except for $\Sigma^0$ and only three of the $J^P=\frac{3}{2}^+$ decuplet baryons are available experimentally \cite{Zyla:2020zbs,Wallace:1995pf, Kotulla:2002cg}. Although all the spatial-ground-state heavy baryons carrying a single charm or bottom quark (except $\Omega_b^{*0}$) have been observed, their magnetic moment has not been measured yet. Increased experimental activity in the heavy flavor sector has escalated the theoretical efforts in the heavy baryon properties, especially, in the case of doubly and triply heavy baryons.
	  
	  Despite the fact that the Standard Model is a well established framework to study interactions of fundamental particles, significant discrepancies can be seen between theoretical predictions and experimental results. However, within the past few decades, several theoretical approaches have been put forth to diminish the inconsistency between theory and experiments. A number of investigations on masses and properties of the heavy charm and bottom baryons based on a variety of models, namely,  naive quark model, non-relativistic and relativistic quark models, hypercentral constituent quark model, Bag Model (BM), Heavy Baryon Chiral Perturbation Theory (HBChPT), pion mean-field approach, Chiral Constituent Quark Model ($\chi$CQM), Light Cone QCD Sum Rule (LCSR),  Lattice QCD (LQCD) and other approaches have been carried out \cite{Choudhury:1976dn,Choudhury:1976dp, Lichtenberg:1976fi,Dattoli:1977nr,Bose:1980vy,Franklin:1981rc,Barik:1984tq,Morpurgo:1989my,Dey:1994qi,Faessler:2006ft,Albertus:2006ya,Roberts:2007ni,Sharma:2010vv,Albertus:2010hi,Bernotas:2012nz,Bernotas:2013eia,Simonis:2018rld,Li:2017cfz,Meng:2017dni,Meng:2018gan,Wang:2018cre,Yang:2018uoj,Yang:2019tst,Yang:2020klp,Aliev:2001ig,Aliev:2001xr,Aliev:2008ay,Aliev:2008sk,Aliev:2009jt,Aliev:2014bma,Aliev:2016xvq,Liu:2009jc,Brown:2014ena,Can:2013tna,Bahtiyar:2018vub,Mathur:2018rwu,Karliner:2014gca,Karliner:2017gml,Karliner:2018hos,Wang:2017kfr,Majethiya:2009vx,Shah:2016vmd,Gandhi:2018lez,Li:2017pxa,Lu:2017meb,Xiabng:2018qsd,Liu:2018euh,Ozdem:2018uue,Ozdem:2019zis,Kumar:2005ei,Dhir:2009ax,Dhir:2013nka}. Furthermore, the radiative decay processes provide an useful mean to study the electromagnetic properties to reveal the inner structure of the heavy baryons, however, the studies on the radiative decays are scarce. Theoretically, radiative decays of heavy flavor baryons have been studied within different approaches \cite{Dey:1994qi,Albertus:2010hi, Bernotas:2013eia,Simonis:2018rld,Wang:2018cre,Yang:2019tst,Aliev:2009jt,Aliev:2014bma,Aliev:2016xvq,Bahtiyar:2018vub,Majethiya:2009vx,Li:2017pxa,Lu:2017meb}. 

	In the present work, we have calculated magnetic and transition moments, and M1 decay widths of ground-state singly, doubly, and triply heavy baryons. Also, we have given the estimates for the magnetic moments in an improved manner by determining hyperfine (one gluon exchange) interaction terms for $s-$, $c-$, and $b-$flavors from precise experimental values of baryon masses within the same flavor sector. Following the Effective Mass Scheme (EMS) \cite{Kumar:2005ei, Dhir:2009ax, Dhir:2013nka}, we have calculated the constituent and effective masses of quarks inside a baryon for both $J^P = \frac{1}{2}^+$ and  $J^P = \frac{3}{2}^+$ baryons. Subsequently, we have calculated the magnetic moments of all the $J^P = \frac{1}{2}^+$ and  $J^P = \frac{3}{2}^+$ baryons up to triply heavy flavors. We believe that the precise experimental information especially in the charm and bottom sectors has helped us to improve our analysis. Furthermore, we have given the predictions for $\frac{3}{2}^{+} \to \frac{1}{2}^{(')+}$ transition moments among singly, doubly, and triply heavy bottom baryons for the first time in EMS. Also, we have calculated the $\frac{1}{2}^{\prime +} \to \frac{1}{2}^{+}$ and $\frac{3}{2}^{+} \to \frac{1}{2}^{(')+}$ transitions in the improved analysis for the singly, doubly, and triply heavy charmed baryons. The major inspiration of the current work is to calculate M1 transition decay widths of heavy flavor baryons in the EMS. In order to make robust predictions and to reduce the uncertainties in radiative M1 decay widths, we have used the theoretically determined masses from LQCD results \cite{Brown:2014ena} in the absence of experimental data. Since the EMS depends mostly on experimental inputs, our results for radiative M1 decay widths are expected to be trustworthy within experimental uncertainties. We have also compared our results with other theoretical models.

The paper is organized in the following way: we give the methodology for EMS in section II. Magnetic moments, transition moments, and M1 radiative decay widths are presented in sections III, IV, and V, respectively. We discuss our results in section VI, and lastly, we give a summary and our conclusions.
	
	\section {Effective quark mass scheme}
	The motive of the EMS lies in the hypothesis that the masses of the quarks inside the baryon are modified as a consequence of one-gluon exchange interaction with the spectator quarks \cite{DeRujula:1975qlm}. In this article, we have calculated the effective quark masses inside the baryons resulting from one-gluon exchange interaction and consequently, observe how it affects the magnetic moments, transition moments, and M1 radiative decay widths. In the EMS, the baryon mass can be written as the sum of the constituent quark masses and the spin-dependent hyperfine interaction among them \cite{Kumar:2005ei, Dhir:2009ax, Dhir:2013nka},
	
	\begin{equation}
	\label{e1}
	{M}_\mathcal{B} = \sum_{i}m_i^\xi = \sum_{i}{m_i} + \sum_{i<j}b_{ij}\bf{s_i.s_j},
	\end{equation}
	where, $m_i^\xi$ represents the effective mass of the quark inside the baryon; $\bf{s_i}$ and $\bf{s_j}$ denote the spin operators of the $i^{th}$ and $j^{th}$ quark, respectively. The $b_{ij}$ for baryons $\mathcal{B}(\it qqq)$, is given by\\
	\begin{equation}
	\label{e2}
	{b}_{ij} =\frac{16\pi\alpha_s}{9m_im_j}<\psi_0|\delta^3(\vec{r})|\psi_0>,
	\end{equation}
	where, $\psi_0$ is the baryon wave function at the origin.\\
	The effects of other kinds of spin-independent interaction terms can be approximated by the renormalization of quark masses. Due to the interaction with other quarks, the mass of the quark inside a baryon gets modified. For (112)- type $J^P = \frac{1}{2}^+$ baryons, we can write\\
	\begin{equation}
	\begin{gathered}
	\label{e3}
	m_1^\xi = m_2^\xi = m + \alpha{b}_{12} + \beta{b}_{13} ,\\
	m_3^\xi = m_3 + 2\beta{b}_{13} ,\\
	\end{gathered}
	\end{equation}
	where, $m_1 = m_2 = m$ and $b_{13} = b_{23}$. The $\alpha$ and $\beta$ parameters are to be determined as follows:\\
	\begin{equation}
	\ M_{\mathcal{B}_{\frac{1}{2}^+}} = 2m + m_3 + \frac{b_{12}}{4} - b_{13},
	\end{equation}
	for
	\begin{equation}
	{\bf{s_1.s_2}} = \frac{1}{4},~ {\bf{{s_1.s_3} = {s_2.s_3}}} = -\frac{1}{2} ,
	\end{equation}
	thus giving 
	\begin{equation}
	\alpha = \frac{1}{8}\mbox{ and }\beta = -\frac{1}{4}.
	\end{equation}
	Therefore, equation (\ref{e1}) may be generalized for $J^P = \frac{1}{2}^+$ baryons as
	\begin{equation}
	\ M_{\mathcal{B}_{\frac{1}{2}^+}} = m_1 + m_2 + m_3 + \frac{b_{12}}{4} - \frac{b_{23}}{2} - \frac{b_{13}}{2}.
	\end{equation}
	By using equation (\ref{e3}), we will obtain more general expressions for effective masses of the quarks inside the baryon, as follows:\\
	
	1. For (112)-type $J^P = \frac{1}{2}^+$ baryons with quarks 1 and 2 being identical,
	\begin{equation}\label{e8}
	m_1^\xi =  m_2^\xi = m + \frac{b_{12}}{8} - \frac{b_{13}}{4},
	\end{equation}
	and
	\begin{equation}
	m_3^\xi= m_3 - \frac{b_{13}}{2}\mbox{ for } 1 = 2\ne3.
	\end{equation}
	
	2. The baryonic states with three different quark flavors (123) can have both anti-symmetric $\Lambda_{[12]3}$-type and symmetric $\Sigma_{\{12\}3}$-type flavor configuration under the exchange of quarks 1 and 2.\\
	
	(a) For $(123)$ $\Lambda$-type, $J^P = \frac{1}{2}^+$ baryons,
	\begin{equation}
	\begin{gathered}
	m_1^\xi = m_1 - \frac{3b_{12}}{8},\\
	m_2^\xi = m_2 - \frac{3b_{21}}{8},
	\end{gathered}
	\end{equation}
	and
	\begin{equation}
	m_3^\xi = m_3 \mbox{ for } 1\ne 2\ne 3.
	\end{equation}
	
	(b) For $(123)$ $\Sigma$-type, $J^P = \frac{1}{2}^+$ baryons,
	\begin{equation}
	\begin{gathered}
	m_1^\xi = m_1 + \frac{b_{12}}{8} - \frac{b_{13}}{4},\\
	m_2^\xi = m_2 + \frac{b_{12}}{8} - \frac{b_{23}}{4},\\
	\end{gathered}
	\end{equation}
	and 
	\begin{equation}\label{e13}
	m_3^\xi = m_3 - \frac{b_{23}}{4} - \frac{b_{13}}{4} \mbox{ for } 1\ne 2\ne 3.
	\end{equation}
	Following the similar procedure described for $J^P = \frac{1}{2}^+$ baryons, the generalized mass formula for different flavor configuration of $J^P = \frac{3}{2}^+$ baryons is given by\\
	\begin{equation}
	\begin{gathered}
	\ M_{{\mathcal{B}}_{\frac{3}{2}^+}} = m_1 + m_2 + m_3 + \frac{b_{12}}{4} + \frac{b_{23}}{4} + \frac{b_{13}}{4},\\
	\mbox{ for }\alpha = \beta = \frac{1}{8}.
	\end{gathered}
	\end{equation}
	Throughout the above discussions 1, 2, 3 represents $\it{u, d, s, c,}$ and $\it{b}$ quarks.\\
	
	1. For (112)-type $J^P = \frac{3}{2}^+$ baryons,
	\begin{equation}\label{e15}
	m_1^\xi =  m_2^\xi = m + \frac{b_{12}}{8} + \frac{b_{13}}{8},
	\end{equation}
	and
	\begin{equation}
	m_3^\xi= m_3 + \frac{b_{13}}{4} \mbox{ for } 1 = 2\ne3.
	\end{equation}
	
	2. For (123)-type $J^P = \frac{3}{2}^+$ baryons,
	\begin{equation}
	\begin{gathered}
	m_1^\xi = m_1 + \frac{b_{12}}{8} + \frac{b_{13}}{8},\\
	m_2^\xi = m_2 + \frac{b_{23}}{8} + \frac{b_{12}}{8},\\
	\end{gathered}
	\end{equation}
	and
	\begin{equation}
	m_3^\xi = m_3 + \frac{b_{13}}{8} + \frac{b_{23}}{8}.
	\end{equation}
	
	3. For (111)-type $J^P = \frac{3}{2}^+$ baryons,
	\begin{equation}
	m_1^\xi = m_2^\xi = m_3^\xi = m + \frac{b_{12}}{4},
	\end{equation}
	and
	\begin{equation}\label{e20}
	b_{12} = b_{23} = b_{13}.
	\end{equation}
	
	The values of constituent quark masses and hyperfine interaction terms $b_{ij} $ are obtained from the experimentally observed baryon masses \cite{Zyla:2020zbs}. We wish to point out that, compared to our previous work \cite{Kumar:2005ei, Dhir:2009ax, Dhir:2013nka} the $b_{ij}$'s, and the effective quark masses are calculated in a more realistic manner for different flavor sectors corresponding to strange, charm, and bottom mass scales. In order to obtain the effective quark masses, especially in the charm and bottom sector, we have calculated the interaction contribution of single gluon exchange term from the corresponding flavor sector. We believe that effective quark masses obtained in this manner will be much more reliable. The numerical values for constituent quark masses and hyperfine interaction terms, with uncertainties arising from the experimental baryon masses \cite{Zyla:2020zbs}, are listed in  Table \ref{Quark mass}.
	
	\begin{table}[!ht]
		\centering
		\captionof{table}{Calculated masses of constituent quarks and hyperfine interaction terms, $b_{ij}( = b_{ji })$ (in MeV).}
		\label{Quark mass}
		\setlength{\tabcolsep}{2pt}
		\begin{tabular}{|c|c||c|c|} 
			\hline
			\textbf{ Quark } & \textbf{Experimental  }&\textbf{Hyperfine Interaction}& \textbf{Experimental }  \\
		
			\textbf{ masses} & \textbf{Inputs \cite{Zyla:2020zbs} }&\textbf{Terms}& \textbf{Inputs \cite{Zyla:2020zbs} }  \\
			\hline\hline
			$m_{u}(= m_{d}) = $ 361.71 $\pm~ 0.33$  & $\Delta^{+}, N $&  $b_{uu}(= b_{ud} = b_{dd})$ = 195.80   $\pm~ 1.30$ & $\Delta^{+}, N $    \\
		
			\textbf{$m_{s}$} = 539.10  $\pm~ 1.20 $ & $\Lambda^{0}$& \textbf{$b_{ss}$} = 74.00 $\pm~ 6.00$  & $\Omega^{-}$\\
			\hline
			\hline
			$m_{c}$ = 1645.70 $\pm~ 3.50$ & $\Omega_{c}^{*0}$ & \textbf{$b_{us}(= b_{ds})$} = 153.20  $\pm~ 0.70$ &  $\Xi_{c}^{*0},\Xi_{c}^{0}$ \\
			&&\textbf{$b_{uc}(= b_{dc})$} = 43.10  $\pm~ 1.60$  &   $\Sigma_{c}^{*+},\Sigma_{c}^{+}$ \\
			
			&&  $b_{sc}$ = 47.10 $\pm~ 1.70$ & $\Omega_{c}^{*0}$, $\Omega_{c}^{0}$ \\

			&&\textbf{$b_{cc}$} = 46.00 $\pm~ 29.00$  &  $\Xi_{cc}^{++}$ \\
			\hline
			\hline
			\textbf{$m_{b}$} = 5043.00  $\pm~ 0.40$  &   $\Lambda_{b}^{0}, N$ 
			&\textbf{$b_{ub}( = b_{db})$} = 13.17 $\pm~ 0.25$ &  $\Sigma_{b}^{*+},\Sigma_{b}^{+}$ \\ 
			&&\textbf{$b_{sb}$} = 16.00  $\pm~ 4.00$ &   $\Xi_{b}^{*0},\Xi_{b}^{0}$ \\
			
			&&\textbf{$b_{bc}$} = 5.04  $\pm~ 0.18$ &  Eq.(\ref{bbc})  \\
			
			&&\textbf{$b_{bb}$} = 1.64 $\pm~ 0.06$ &   Eq.(\ref{bbb})  \\
			\hline
		\end{tabular}
	\end{table}
	
	We wish to remark here that in contrast to our previous work, for doubly and triply heavy baryons, the hyperfine interaction terms are calculated from known experimental values. For the first time we have estimated $b_{cb}$ and $b_{bb}$ from the hyperfine interaction terms $b_{sc}$ and $b_{sb}$ that are obtained from the experimentally known masses. We use the symmetry relations \cite{Kumar:2005ei, Dhir:2009ax, Dhir:2013nka} to get,
	\begin{equation} \label{bbc}
	\begin{gathered}
	b_{cb} =\left( \frac{m_{s} }{m_{b}} \right) b_{sc} = 5.04\pm 0.18~{\rm MeV}, \mbox{ and }
	b_{cb} = \left(\frac{m_{s} }{m_{c}} \right) b_{sb} = 5.24 \pm 1.30~{\rm MeV},\\ 
	\end{gathered} 
	\end{equation}
	since both the values are roughly same, we use
	\begin{equation}
	b_{cb} = 5.04\pm 0.18 ~ {\rm MeV}.\\ \non
	\end{equation}
	Furthermore, we get
	\begin{equation}\label{bbb}
	\begin{gathered}
	b_{bb} =\left( \frac{m_{s} m_{c}}{m_{b}^2}\right) b_{sc} = 1.64\pm0.06~{\rm MeV}, \mbox{ and } b_{bb} = \left( \frac{m_{s} m_{b}}{m_{b}^2}\right) b_{sb} = 1.70\pm0.40~{\rm MeV}, 
	\end{gathered} 
	\end{equation}
we use,
 $$b_{bb} = 1.64~\pm~0.06 ~\text{MeV}.$$
It may be noted that the numerical uncertainties in the masses and hyperfine interaction terms are calculated from  quadrature approach assuming normal distribution. Since, the uncertainties in the experimentally measured baryon masses are very small, we anticipate smaller percentage  errors in the numerical calculations. Percentage error in all the parameters (as shown in Table \ref{Quark mass}) are less than $7\%$ except for $b_{cc}$ and $b_{sb}$, for which we get $63\%$ and $25\%$, respectively. The larger uncertainties in $b_{cc}$ and $b_{sb}$ parameters are propagating from the uncertainties of baryon masses, constituent quark masses, and other hyperfine interaction terms used to calculate these parameters. However, these large uncertainties in the above-mentioned hyperfine interaction terms are absorbed in the effective quark masses   owing to their smaller magnitude, especially in charm and bottom sectors. Therefore, the propagation of uncertainties in magnetic moments, transition moments and radiative decays are expected to be small.

	\subsection{Effective quark masses for $J^{P}   =  \frac{1}{2}^{+}$  baryons } 
	Using the Eqs. ($\ref{e8}$) - ($\ref{e13}$) and inputs from  Table \ref{Quark mass}, we obtain the effective masses of quarks inside the $J^{P} = \frac{1}{2}^{+} $ baryons as given below.
	\begin{enumerate}
		
	\item  For singly heavy baryons,
	
	\[m_{u}^{\Lambda _{c} } =m_{d}^{\Lambda _{c} } = 288.3\pm0.6 {~\rm MeV}, m_{c}^{\Lambda _{c} }=  1645.7\pm3.5 ~\rm MeV;\]
	
	\[m_{u}^{\Sigma _{c} } =m_{d}^{\Sigma _{c} } = 375.4\pm0.5 {~\rm MeV}, m_{c}^{\Sigma _{c} } =1624.2\pm3.5{~\rm MeV};\]
	
	\[m_{u}^{\Xi^{'} _{c} } =m_{d}^{\Xi^{'} _{c} } = 370.1\pm0.5 {~\rm MeV}, m_{s}^{\Xi^{'} _{c} } = 546.5\pm1.3 {~\rm MeV}, m_{c}^{\Xi^{'} _{c} } = 1623.2\pm3.5 {~\rm MeV};\]
	
    \[m_{u}^{\Xi_{c} } =m_{d}^{\Xi_{c} } = 304.3\pm 0.4{~\rm MeV}, m_{s}^{\Xi_{c} }  = 481.7\pm1.2 {~\rm MeV}, m_{c}^{\Xi_{c} } = 1645.7\pm3.5 {~\rm MeV};\]
    
    \[m_{s}^{\Omega _{c} } = 536.6\pm1.4 {~\rm MeV},m_{c}^{\Omega _{c} } = 1622.0\pm4.0 {~\rm MeV};\]
    
	\[m_{u}^{\Lambda _{b} } =m_{d}^{\Lambda _{b} } = 288.3\pm0.6 {~\rm MeV}, m_{b}^{\Lambda _{b} }=  5043.0\pm0.4 ~\rm MeV;\]
	
	\[m_{u}^{\Xi_{b} } =m_{d}^{\Xi_{b} } = 304.3\pm 0.4{~\rm MeV}, m_{s}^{\Xi_{b} }  = 481.7\pm1.2 {~\rm MeV}, m_{b}^{\Xi_{b} } = 5043.0\pm 0.4{~\rm MeV};\] 
	                     
	\[m_{s}^{\Omega _{b} } = 544.4\pm 1.7{~\rm MeV},m_{b}^{\Omega _{b} } = 5035.0\pm2.0 {~\rm MeV};\]
	
	\[m_{u}^{\Sigma _{b} } = m_{d}^{\Sigma _{b} } = 382.9\pm0.4 {~\rm MeV}, m_{b}^{\Sigma _{b} } = 5036.4\pm0.4 {~\rm MeV};\]
	
	\[m_{u}^{\Xi^{'} _{b} } = m_{d}^{\Xi^{'} _{b} } = 377.6\pm0.4 {~\rm MeV}, m_{s}^{\Xi^{'} _{b} } = 554.3\pm 1.6{~\rm MeV}, m_{b}^{\Xi^{'} _{b} } = 5035.7\pm1.1{~\rm MeV}.\]
	
	\item  For doubly heavy baryons,
	
	\[m_{u}^{\Xi _{_{cc} } } = m_{d}^{\Xi _{_{cc} } } = 340.2\pm0.9 {~\rm MeV},m_{c}^{\Xi _{_{cc} } } = 1641.0\pm5.0 {~\rm MeV};\]
	
	\[m_{s}^{\Omega _{cc} } = 515.6\pm 1.5{~\rm MeV}, m_{c}^{\Omega _{cc} } = 1640.0\pm 5.0{~\rm MeV};\]
	
	\[m_{u}^{\Xi_{cb}  } = m_{d}^{\Xi_{cb}  } = 345.5\pm 0.7{~\rm MeV}, m_{c}^{\Xi_{cb} } = 1630.0\pm4.0 {~\rm MeV},m_{b}^{\Xi_{cb} } = 5043.0\pm0.4 {~\rm MeV};\]
	
	\[m_{s}^{\Omega_{cb} } = 521.4\pm 1.4{~\rm MeV}, m_{c}^{\Omega_{cb} } = 1628.0\pm4.0 {~\rm MeV}, m_{b}^{\Omega_{cb} } = 5043.0\pm0.4 {~\rm MeV};\]
	
	\[m_{u}^{\Xi^{'} _{cb}} = m_{d}^{\Xi^{'} _{cb}  } = 363.8\pm0.4 {~\rm MeV}, m_{c}^{\Xi^{'} _{cb}  } = 1649.8\pm3.5 {~\rm MeV}, m_{b}^{\Xi^{'} _{cb}  } = 5038.4\pm0.4 {~\rm MeV}; \]
	
	\[m_{s}^{\Omega _{cb}^{'} } = 541.0\pm1.6 {~\rm MeV}, m_{c}^{\Omega _{cb}^{'} } = 1650.3\pm3.5 {~\rm MeV}, m_{b}^{\Omega _{cb}^{'} } = 5037.7\pm1.1 {~\rm MeV};\]
	
	\[m_{u}^{\Xi _{_{bb} } } = m_{d}^{\Xi _{_{bb} } } = 355.1\pm0.4 {~\rm MeV}, m_{b}^{\Xi _{_{bb} } } = 5039.9\pm0.4 {~\rm MeV};\]
	
	\[m_{s}^{\Omega _{bb} } = 531.1\pm2.3 {~\rm MeV}, m_{b}^{\Omega _{bb} } = 5039.2\pm1.1{~\rm MeV}.\]
	
	\item  For triply heavy baryons,
	
	\[ m_{c}^{\Omega _{ccb} } = 1650.0\pm5.0{~\rm MeV}, m_{b}^{\Omega _{ccb} } = 5040.5\pm0.4 {~\rm MeV};\]
	
	\[m_{c}^{\Omega _{cbb} } = 1643.2\pm3.5 {~\rm MeV}; m_{b}^{\Omega _{cbb} } = 5041.9\pm0.4 {~\rm MeV}.\]
	 
	\end{enumerate}	
	\subsection{Effective quark masses for $J^{P}   =  \frac{3}{2}^{+}$ baryons } 
	For $J^{P} = \frac{3}{2}^{+}$ baryons, we use the Eqs. (\ref{e15}) - (\ref{e20}) and inputs from Table \ref{Quark mass} to obtain the effective masses of quarks as given below.
	\begin{enumerate}
		
		\item  For singly heavy baryons,
	\[ m_{u}^{\Sigma _{_{c} }^{*} } = m_{d}^{\Sigma _{_{c} }^{*} } = 391.6\pm0.4 {~\rm MeV}, m_{c}^{\Sigma _{_{c} }^{*} } = 1656.5\pm 3.5{~\rm MeV};\]
	
	\[ m_{u}^{\Xi_{c}^{*} } = m_{d}^{\Xi _{c}^{*} } = 386.2\pm0.4 {~\rm MeV}, m_{s}^{\Xi_{c}^{*} } = 564.1\pm1.2 {~\rm MeV}, m_{c}^{\Xi _{c}^{*} } = 1657.0\pm3.5 {~\rm MeV};\]
	
	\[ m_{s}^{\Omega _{c} ^{*}} = 554.2\pm1.4 {~\rm MeV}, m_{c}^{\Omega _{c}^{*} } = 1657.5\pm3.5 {~\rm MeV};\]
	
		\[ m_{u}^{\Sigma _{_{b} }^{*} } = m_{d}^{\Sigma _{_{b} }^{*} } = 387.8\pm0.4 {~\rm MeV}, m_{b}^{\Sigma _{_{b} }^{*} } = 5046.3\pm0.4 {~\rm MeV};\]
		
		\[ m_{u}^{\Xi_{b}^{*} } = m_{d}^{\Xi _{b}^{*} } = 382.5\pm 0.3{~\rm MeV}, m_{s}^{\Xi_{b}^{*} } = 560.3\pm1.3 {~\rm MeV}, m_{b}^{\Xi _{b}^{*} } = 5046.6\pm0.6{~\rm MeV};\]
		
		\[ m_{s}^{\Omega _{b} ^{*}} = 550.4\pm1.4 {~\rm MeV}, m_{b}^{\Omega _{b}^{*} } = 5047.0\pm1.1 {~\rm MeV}.\]
		
		 \item  For doubly heavy baryons,
	\[  m_{u}^{\Xi _{cc}^{*} } =  m_{d}^{\Xi _{cc}^{*} } = 372.5\pm0.5 {~\rm MeV}, m_{c}^{\Xi _{cc}^{*} } = 1657.0\pm 5.0{~\rm MeV};\] 
	
	\[  m_{s}^{\Omega _{cc}^{*} } = 550.9\pm 1.3{~\rm MeV}, m_{c}^{\Omega _{cc}^{*} } = 1657.0\pm5.0 {~\rm MeV};\]
	
		\[ m_{u}^{\Xi _{cb}^{*} } =  m_{d}^{\Xi _{cb}^{*} } = 368.7\pm0.4 {~\rm MeV}, m_{c}^{\Xi _{cb}^{*} } = 1651.7\pm3.5 {~\rm MeV}, m_{b}^{\Xi _{cb}^{*} } = 5045.3\pm0.4 {~\rm MeV};\]
		
		\[  m_{s}^{\Omega _{cb}^{*} } = 547.0\pm1.3 {~\rm MeV}, m_{c}^{\Omega _{cb}^{*} } = 1652.2\pm3.5 {~\rm MeV}, m_{b}^{\Omega _{cb}^{*} } = 5045.6\pm0.6 {~\rm MeV};\]
		
		\[  m_{u}^{\Xi _{bb}^{*} } = m_{d}^{\Xi _{bb}^{*} } = 365.0\pm0.3 {~\rm MeV}, m_{b}^{\Xi _{bb}^{*} } = 5044.9\pm0.4 {~\rm MeV}; \] 
		           
		\[ m_{s}^{\Omega _{bb}^{*} } = 543.1\pm 1.6{~\rm MeV}, m_{b}^{\Omega _{bb}^{*} } = 5045.2\pm0.6{~\rm MeV}.\]
		
		\item  For triply heavy baryons,
	\[ m_{c}^{\Omega _{ccc}^{*} } = 1657.0\pm8.0 {~\rm MeV};\]
	
		\[ m_{c}^{\Omega _{_{ccb} }^{*} } = 1652.0\pm5.0{~\rm MeV}, m_{b}^{\Omega _{_{ccb} }^{*} } = 5044.3\pm 0.4{~\rm MeV};\]
		
		\[ m_{c}^{\Omega _{cbb}^{*} } = 1647.0\pm 3.5{~\rm MeV}, m_{b}^{\Omega _{cbb} ^{*} } = 5043.8\pm0.4 {~\rm MeV};   \]
		
		\[m_{b}^{\Omega _{bbb}^{*} } = 5043.4\pm0.4 {~\rm MeV}.\] 
		
	\end{enumerate}

	\section{ Magnetic Moments of  $\pmb{(J^{P}   =  \frac{1}{2}^{+} )}$ and  $\pmb{(J^{P}   =  \frac{3}{2}^{+} )}$ Baryons}
	
	In the EMS, magnetic moments of $J^{P}   =  \frac{1}{2}^{+} $ baryons are obtained by sandwiching the following magnetic moment operator between the appropriate baryon wave functions: 
	
	\begin{equation} \label{e23} 
	\pmb{\mu }=\sum _{i}\mu _{i}^{\xi} \pmb{\sigma }_{i},                           \end{equation} 
	where, $\pmb{\sigma }_{i}$ is the Pauli's spin matrix and $\mu _{i}^{\xi}$, the effective quark magnetic moment defined as
	\begin{equation}
	\mu _{i}^{\xi} =\frac{e_{i} }{2m_{i}^{\xi}},
	\end{equation}    
	where \textit{i = u, d, s, c}, and \textit{b}; $e_i$ represent the respective quark charges.
	\par The magnetic moment of a baryon depends on its constituent quark flavors and on their spin configuration, and can thus be written as,
	
	\begin{equation}
	\mu(\mathcal{B})=  \langle{\mathcal{B}
	}|\bm{\mu}|{\mathcal{B}}\rangle
	\end{equation}
	where, $\mathcal{B}$ = $\mathcal{B}{\it(q_{1}q_{2}q_{3})}$ denotes the spin-flavor wavefunction of the baryons with $\it(q_{1}q_{2}q_{3})$ quarks composition. The general expressions for flavor degenerate octet baryon magnetic moments are given by:
	\begin{equation}
	\begin{gathered}
	\mu (\mathcal{B}) = \frac{1}{3}(4\mu^\mathcal{\xi} _{1} - \mu^\mathcal{\xi}_{2}),\\
	\mu (\mathcal{B}) =\mu^\mathcal{\xi} _{3},  \\
	\mu (\mathcal{B^{\prime }}) = \frac{1}{3}(2\mu^\mathcal{\xi} _{1}+2\mu^\mathcal{\xi}_{2}-\mu ^\mathcal{\xi}_{3}),
	\end{gathered}
	\end{equation}
	where, $\mu^{\mathcal{\xi}} _{i}$ denote the effective magnetic moments of first, second and third quarks, respectively. 
	
	Proceeding in a way similar to $J^P=\frac{1}{2}^+$, the magnetic moments of decuplet  $J^P=\frac{3}{2}^+$ baryons are obtained by sandwiching the same magnetic moment operator (\ref{e23}) between the corresponding baryon wave functions. The general expression for decuplet baryon magnetic moments is given by,
	\begin{align}
	\mu (\mathcal{B^{*}}) & =\mu^\mathcal{\xi} _{1}+\mu^\mathcal{\xi} _{2}+\mu ^\mathcal{\xi}_{3}.
	\end{align}
	The wave functions of ground state baryons, which are used in the above equations, can be constructed by coupling the spins of the first two quarks to an intermediate spin state S and then adding the third one to form the baryon with the resulting spin $J$. We adopt the convention that $\left[q_{1} q_{2}\right]$ denotes anti-symmetric (S = 0) and $\left\{q_{1} q_{2}\right\}$ denote symmetric (S = 1) combinations of quark flavor indices (with respect to the interchange
	of $q_1$ and  $q_2$):
	
	\begin{equation}
	\begin{aligned}
	|\mathcal{B}\rangle &=\left|\left[q_{1} q_{2}\right]^{S = 0} q_{3}, J=\frac{1}{2}\right\rangle \\
	\left|\mathcal{B}^{\prime}\right\rangle &=\left|\left\{q_{1} q_{2}\right\}^{S = 1} q_{3}, J=\frac{1}{2}\right\rangle \\
	\left|\mathcal{B}^{*}\right\rangle &=\left|\left\{q_{1} q_{2}\right\}^{S = 1} q_{3}, J=\frac{3}{2}\right\rangle .
	\end{aligned}
	\end{equation}
	The same notation has been kept throughout this article. Numerical results for the magnetic moments of $J^{P} = \frac{1}{2}^{+} $ and  $J^{P}   =  \frac{3}{2}^{+} $ baryons using effective quark masses obtained in  the last section are listed in Tables \ref{t1}, \ref{t2}, \ref{t3}, and \ref{t4}, respectively. Aforementioned, due to higher precision and small uncertainties in experimentally measured masses, the uncertainties propagating to magnetic and transition moments are less than $4\%$. Therefore, the results are given with upto  four significant digits. Also, we have given the theoretical predictions from other models in the respective Tables for comparison. 

\section{Transition Moments Relations}
	
	Proceeding in the same manner as in the last section, we calculate the transition moments of $\frac{1}{2}^{'+} \rightarrow \frac{1}{2}^{+}$, $\frac{3}{2}^{+} \rightarrow \frac{1}{2}^{+}$, and $\frac{3}{2}^{+} \rightarrow \frac{1}{2}^{'+}$ by sandwiching the magnetic moment operator defined by equation (\ref{e23}) between the initial and final state $\frac{3}{2}^{+}$, $\frac{1}{2}^{+}$, and $\frac{1}{2}^{'+}$ baryon wave functions. The general expressions for these transition moments are given by: 
	\begin{equation}
	\begin{gathered}
	\begin{aligned}
	\mu_{\frac{1}{2}^{'+}  \to \frac{1}{2}^{+}} & = \sqrt{1\over3}\left[\mu^{\xi}(2)-\mu^{\xi}(1)\right],\\
	\mu_{\frac{3}{2}^{+}  \to \frac{1}{2}^{+}}  & = \sqrt{2\over3} \left[\mu^{\xi}(1)-\mu^{\xi}(2)\right], \\
	\mu_{\frac{3}{2}^{+}  \to \frac{1}{2}^{'+}}  & = {\sqrt{2}\over3}\left[\mu^{\xi}(1)+\mu^{\xi}(2)-2\mu^{\xi}(3)\right].\\
	\end{aligned}
	\end{gathered}
	\end{equation} 
	
	Expressions for $\mu_{\frac{3}{2}^{+}  \to \frac{1}{2}^{+}}$ transition magnetic moments (in $\mu_N$) of charmed baryons are listed in \cite{Dhir:2009ax}. The expressions for transition moments of $\mu_{\frac{1}{2}^{'+}  \to \frac{1}{2}^{+}}$ and $\mu_{\frac{3}{2}^{+}  \to \frac{1}{2}^{(')+}}$ singly, doubly, and triply heavy baryons are listed in Table \ref{t5} and \ref{t6}, respectively. To evaluate $\mu_{\frac{1}{2}^{'+}  \to \frac{1}{2}^{+}}$ and $\mu_{\frac{3}{2}^{+}  \to \frac{1}{2}^{(')+}}$  transition moments, we take the geometric mean of effective quark masses of the constituent quarks of initial and final state baryons,
	\begin{equation}
	\begin{gathered} \label{gm}
m_i^{\xi}({\mathcal B_J^{\prime (*)}} \to \mathcal{B}_{J})=\sqrt{m_i^{\xi}({\mathcal B_J^{\prime (*)}}) m_i^{\xi}({\mathcal B_{J}})},\\
	\end{gathered}
	\end{equation}
	where, symbols have their usual meaning.
	Using the Eqs. ($\ref{e8}$) - ($\ref{e13}$) and ($\ref{e15}$) - ($\ref{e20}$), we calculate the transition ($\tbt \to \obt^{(\prime)}$)\footnote[4]{the masses for $\obt \to \obtp$ can be obtained in a similar manner using Eq. (\ref{gm}).} masses of baryons as follows:
	
	\begin{enumerate}
		
		\item  For singly heavy charmed baryons, 
		
		\[m_{u}^{\Lambda _{c}} = m_{d}^{\Lambda _{c}} = 336.0\pm0.4 {~\rm MeV}, m_{c}^{\Lambda _{c} }= 1651.1\pm2.5 ~\rm MeV;\]
		
		\[m_{u}^{\Xi_{c} } = m_{d}^{\Xi_{c} } = 342.8\pm 0.3{~\rm MeV}, m_{s}^{\Xi_{c} }  = 521.3\pm0.9 {~\rm MeV}, m_{c}^{\Xi_{c} } = 1651.3\pm 2.5{~\rm MeV};\]   
		
		\[m_{u}^{\Sigma _{c} } = m_{d}^{\Sigma _{c} } = 383.4\pm 0.3{~\rm MeV}, m_{c}^{\Sigma _{c} } = 1640.2\pm2.5 {~\rm MeV};\]
		
		\[m_{u}^{\Xi^{'} _{c} } = m_{d}^{\Xi^{'} _{c} } = 378.1\pm0.3 {~\rm MeV}, m_{s}^{\Xi^{'} _{c} } = 555.2\pm0.9 {~\rm MeV}, m_{c}^{\Xi^{'} _{c} } = 1640.0\pm2.5 {~\rm MeV};\]
		
		\[m_{s}^{\Omega _{c} } = 545.3\pm1.0{~\rm MeV},m_{c}^{\Omega _{c} } = 1639.7\pm 2.5{~\rm MeV}.\]
		
		\item  For doubly heavy charmed baryons,
		
		\[m_{u}^{\Xi_{cc}  } = m_{d}^{\Xi_{cc}  } = 356.0\pm0.5 {~\rm MeV}, m_{c}^{\Xi_{cc} } = 1649.0\pm4.0 {~\rm MeV};\]
		
		\[m_{s}^{\Omega_{cc} } = 532.9\pm1.0{~\rm MeV}, m_{c}^{\Omega_{cc} } = 1648.0\pm4.0 {~\rm MeV}.\]
		
		\item  For singly heavy bottom baryons, 
		
		\[m_{u}^{\Lambda _{b} } = m_{d}^{\Lambda _{b} } = 334.4\pm0.4 {~\rm MeV}, m_{b}^{\Lambda _{b} }= 5044.6\pm0.3 ~\rm MeV;\]
		
		\[m_{u}^{\Xi_{b} } = m_{d}^{\Xi_{b} } = 341.2\pm0.3 {~\rm MeV}, m_{s}^{\Xi_{b} }  = 519.5\pm0.9 {~\rm MeV}, m_{b}^{\Xi_{b} } = 5044.8\pm0.4 {~\rm MeV};\]
		
		\[m_{s}^{\Omega _{b} } = 547.3\pm1.1 {~\rm MeV},m_{b}^{\Omega _{b} } = 5041.0\pm1.1 {~\rm MeV};\]
		
		\[m_{u}^{\Sigma _{b} } = m_{d}^{\Sigma _{b} } = 385.4\pm 0.3{~\rm MeV}, m_{b}^{\Sigma _{b} } = 5041.4\pm0.3 {~\rm MeV};\]
		
		\[m_{u}^{\Xi^{'} _{b} } = m_{d}^{\Xi^{'} _{b} } = 380.0\pm 0.2{~\rm MeV}, m_{s}^{\Xi^{'} _{b} } = 557.2\pm1.0 {~\rm MeV}, m_{b}^{\Xi^{'} _{b} } = 5041.2\pm 0.6{~\rm MeV}.\]
		
		\item  For doubly heavy baryons,
		
		\[m_{u}^{\Xi_{cb}  } = m_{d}^{\Xi_{cb}  } = 357.0\pm0.4 {~\rm MeV}, m_{c}^{\Xi_{cb} } = 1640.6\pm2.5 {~\rm MeV},m_{b}^{\Xi_{cb} } = 5044.1\pm0.3 {~\rm MeV};\]
		
		\[m_{s}^{\Omega_{cb} } = 534.1\pm0.9 {~\rm MeV}, m_{c}^{\Omega_{cb} } = 1640.1\pm 2.5{~\rm MeV}, m_{b}^{\Omega_{cb} } = 5044.3\pm 0.4{~\rm MeV};\]
		
		\[m_{u}^{\Xi^{'} _{cb}} = m_{d}^{\Xi^{'} _{cb}  } = 366.3\pm0.3 {~\rm MeV}, m_{c}^{\Xi^{'} _{cb}  } = 1650.8\pm2.5 {~\rm MeV}, m_{b}^{\Xi^{'} _{cb}  } = 5041.9\pm0.3 {~\rm MeV}; \]
		
		\[m_{s}^{\Omega _{cb}^{'} } = 544.0\pm 1.0{~\rm MeV}, m_{c}^{\Omega _{cb}^{'} } = 1651.3\pm2.5 {~\rm MeV}, m_{b}^{\Omega _{cb}^{'} } = 5041.7\pm0.6 {~\rm MeV};\]
		
		\[m_{u}^{\Xi _{_{bb} } } = m_{d}^{\Xi _{_{bb} } } = 360.0\pm0.2 {~\rm MeV}, m_{b}^{\Xi _{_{bb} } } = 5042.4\pm0.3 {~\rm MeV};\]
		
		\[m_{s}^{\Omega _{bb} } = 537.1\pm1.4 {~\rm MeV}, m_{b}^{\Omega _{bb} } = 5042.2\pm0.6 {~\rm MeV}.\]
		
		\item  For triply heavy baryons,
		
		\[ m_{c}^{\Omega _{ccb} } = 1651.0\pm4.0 {~\rm MeV}, m_{b}^{\Omega _{ccb} } = 5042.4\pm0.3 {~\rm MeV};\]
		
		\[m_{c}^{\Omega _{cbb} } = 1645.1\pm2.5 {~\rm MeV}; m_{b}^{\Omega _{cbb} } = 5042.9\pm0.3 {~\rm MeV}.\] 
		
	\end{enumerate}
	The numerical results for the transition moments of heavy flavor baryons with uncertainties are presented in Tables \ref{t7}, \ref{t8}, and \ref{t9}. As expected, the uncertainties in transition moments are less than $4\%$.
	
	\section{RADIATIVE DECAY WIDTHS}
	The EM structure of the baryon decuplet is more complicated than that of the baryon octet since all the decuplet baryons have spin$-\tbt$. Thus, a decuplet baryon has two more terms of the EM form factors than the baryon octet, \textit{i.e.}, the electric quadrupole (E2) form factors and the magnetic octupole (M3) form factors in addition to the electric monopole (E0) and magnetic dipole (M1) form factors. The experimental determination of the radiative decay width of heavy baryons is important for the understanding of its intrinsic properties. We would like to continue our query with analysis for M1 partial widths of the ground state heavy baryons. We ignore the transition of type E2, which is expected to be much smaller in magnitude \cite{Bahtiyar:2018vub, Yang:2019tst}, when compared to M1. The radiative decay widths of the decay type $\mathcal B_J^{\prime (*)}\to \mathcal{B}_J\gamma$ (Ref. \cite{Dey:1994qi, Simonis:2018rld}) is given by
	\begin{equation}
	\Gamma(\mathcal B_J^{\prime (*)}\to \mathcal{B}_J\gamma) = {\alpha\omega^3\over m^2_p}{2\over (2J+1)}|\mu(\mathcal B_J^{\prime (*)}\to \mathcal{B}_J)|^2,
	\end{equation}
	where, \begin{equation}
	\omega = \frac{M^2_{\mathcal{B}^{'(*)}} - M^2_\mathcal{B} } {2M_{\mathcal{B}^{'(*)}}},
	\end{equation}
	is the photon momentum in the center-of-mass system of the initial baryon states.
	Here, $\mu(\mathcal B_J^{\prime (*)}\to \mathcal{B}_J)$ is the transition moments (in $\mu_N$), J is the spin quantum number for parent state, and $M_{\mathcal{B^{'(*)}}}$ and $M_\mathcal{B}$ are the masses of initial and final baryon state, respectively. The radiative decay widths depend on the nature of the interaction that causes the decay as well as on the properties of the initial and final baryon states involved; therefore, are quantity of central interest in the present analysis.
	
	Here, we use already calculated transition moments (in the last section) for the evaluation of spin $\frac{1}{2}^{'+}\rightarrow\frac{1}{2}^{+}$ and $\frac{3}{2}^{+}\rightarrow\frac{1}{2}^{(')+}$ decay widths. As already pointed out by Bernotas \textit{et al.} \cite{Bernotas:2013eia, Simonis:2018rld}, these transitions are affected by uncertainties arising from the absence of experimental masses and evaluation of photon momentum. In order to give reliable predictions, we have used the experimental masses of baryons \cite{Zyla:2020zbs} and LQCD estimates for the unobserved baryon masses \cite{Brown:2014ena}. The uncertainties arising from the baryon masses in evaluation of photon momenta are less than $5\%$ as shown in Tables \ref{t10} and \ref{t11}. We compare $\omega$ used in our EMS with predictions from BM \cite{Bernotas:2013eia}. The calculated radiative decay widths for singly, doubly, and triply heavy baryons, with uncertainties from transition moments and photon momenta are listed in Tables \ref{t12}, \ref{t13}, and \ref{t14}. The uncertainties in these results range up to $14\%$. 

	\section{Numerical Results and Discussion}
	
	In the present work, we have focused on transition moments, and M1 decay widths of ground-state singly, doubly, and triply heavy baryons. We have also given the estimates for the magnetic moments of these decays. We have calculated the one gluon exchange interaction terms from the known experimental masses for heavy flavors, which are expected to be more reliable. One of the key features of EMS is to treat all the quarks at the same footing making it virtually parameter independent (as $b_{ij}$'s are fixed from experimental values). Following the aforementioned approach, we have calculated the constituent and effective masses of quarks inside a baryon by using experimentally observed masses for both $J^P = \frac{1}{2}^+$ and  $J^P = \frac{3}{2}^+$ baryons in Section II. We have included the experimental uncertainties in the evaluation of constituent quark masses and hyperfine interaction terms. We then proceed to calculate the magnetic moments of all the $J^P = \frac{1}{2}^+$ and  $J^P = \frac{3}{2}^+$ baryons up to $b = 3$, we believe that the precise experimental information especially in charm and bottom sectors has helped us to improve our analysis. 
	
	In this paper, we give the predictions for $\frac{3}{2}^{+} \to \frac{1}{2}^{+}$ transition moments involving bottom flavor for the first time in EMS. Also, we reproduce all the  $\frac{1}{2}^{\prime +} \to \frac{1}{2}^{+}$ and $\frac{3}{2}^{+} \to \frac{1}{2}^{+}$ transitions in the improved analysis in the charmed baryons. One of the main objectives of the current work is to calculate M1 transition decay widths for both charm and bottom baryons. As mentioned before, to reduce the uncertainties in the calculation of radiative M1 decay widths, the experimental values of the baryon masses (wherever available) are used. However, we adopt the theoretical estimates from LQCD results \cite{Brown:2014ena}, when experimental data are absent. This way we have tried to obtain accurate values for the photon momenta with uncertainties ranging to a maximum of 5\%. Since the EMS depends mostly on experimental inputs, our results for radiative M1 decay widths are expected to be trustworthy within experimental uncertainties. We wish to remark that for the sake of comparison, we have given the results from the other models for the magnetic moments in the corresponding Tables. Although our results have improved, the conclusions from the previous work \cite{Kumar:2005ei, Dhir:2009ax, Dhir:2013nka} on magnetic moments remain valid. In this work, we will focus our discussions on transition moments and decay widths in the following subsections.
	
	\subsection{Transition Moments}
	We wish to emphasize that in our analysis we have taken the magnitude of transition moments with signs. The sign difference in various models can be solely attributed to the phase convention adopted, which does not affect the physical process in any way. 
	
	\subsection* {Spin $ (\frac{1}{2}^{\prime \,+} \rightarrow  \frac{1}{2}^+)$ transition moments}
	
	\begin{enumerate}
		
		\item[i.] Our results of transition moments involving singly heavy charmed baryons $\Sigma_c^{+}\rightarrow \Lambda_c^{+}$ and $\Xi_c^{\prime\,+}\rightarrow \Xi_c^{+}$ compare well with other theoretical models \cite{Simonis:2018rld, Yang:2019tst, Sharma:2010vv}, except for $\Xi_c^{\prime \,0}\rightarrow \Xi_c^{0}$ transition. Our results for $\Xi_c^{\prime \,0}\rightarrow \Xi_c^{0}$ are consistent with those from the HBChPT \cite{Wang:2018cre} and pion mean-field approach \cite{Yang:2019tst}. 
		
		\item[ii.] For singly heavy bottom baryons, our results are in agreement with the predictions of BM \cite {Simonis:2018rld}, HBChPT \cite{Wang:2018cre}, and chiral quark-soliton model ($\chi$QSM) \cite{Yang:2019tst} with few exceptions. However, the theoretical predictions for doubly heavy baryons exist scarcely.
		We compare our results with BM \cite{Bernotas:2012nz, Simonis:2018rld} which are small as compared to our results. We wish to remark here that, discrepancies between various models may arise due to different inputs and parameter dependence.  In addition, some authors have used mixing between anti-symmetric and symmetric states of the same flavor which is expected to be small and can be estimated reliably from precise experimental masses.
		
	\end{enumerate}
	
	\subsection* {Spin $ (\frac{3}{2}^+ \rightarrow  \frac{1}{2}^+)$ transition moments}
	\begin{enumerate}
		
		\item[i.] For singly charmed baryons, our predictions of transition moments are comparable with those from the $\chi$CQM \cite{Sharma:2010vv}, BM \cite{Simonis:2018rld}, HBChPT \cite{Wang:2018cre}, and  $\chi$QSM  \cite{Yang:2019tst}, except for the $\Sigma_c^{*} \rightarrow \Sigma_c$, $\Xi_c^{*} \rightarrow \Xi_c^{(\prime )}$ and $\Omega_{c}^{*0} \to \Omega_{c}^0$ transitions. The significant deviation in results may appear due to mixing effects, which have been ignored in the present work. 
		
		\item[ii.] Results from the HBChPT \cite{Wang:2018cre}, $\chi$QSM \cite{Yang:2019tst}, and LCSR \cite{Aliev:2009jt} for the charmed baryons differ from our results. However, it is to be kept in mind that these results being model dependent differ widely from our assumptions. Furthermore, we have adopted a reasonably accurate analysis and have partially taken into account the symmetry breaking effects, through masses, which can significantly impact values for the transition moments. 
		
		\item[iii.] Considering the results for doubly charmed baryons, we have compared our results with those from the $\chi$CQM \cite{Sharma:2010vv}, BM \cite{Simonis:2018rld}, and HBChPT \cite{Li:2017pxa}. Our results are consistent with the BM \cite{Simonis:2018rld} predictions, but are smaller than chiral perturbation theory expectations \cite{Li:2017pxa}.
		
		\item[iv.] For the singly bottom baryon transitions, our results are close to the predictions of BM \cite{Simonis:2018rld}, HBChPT \cite{Wang:2018cre}, and $\chi$QSM  \cite{Yang:2019tst}, with some exceptions. However, for the doubly heavy bottom baryons, our results are in better agreement with the BM \cite{Simonis:2018rld}, except for the transition involving $\Xi _{cb}^{\prime \,0}$ in the final state. Here also, the disagreement may be attributed to the state mixing effects in these states.
		
		\item[v.] In the triply heavy sector, transition moments for $\Omega _{ccb}^{*+}\rightarrow \Omega _{ccb}^{+}$ and  $\Omega _{cbb}^{*0}\rightarrow \Omega _{cbb}^{0}$ are in good agreement with BM \cite{Simonis:2018rld}.
	\end{enumerate}
	In general, the consistency between the final results of a variety of approaches is evident. It can be argued that for the static properties of baryons, the non-relativistic constituent quark model approach is completely equivalent to a parameterization of the relativistic field theory of strong interactions in a spin-flavor basis \cite{Morpurgo:1989my,Dillon:1995qw}. So, the use of the one-body operator is justified in light of the decoupling of spatial and spins of the ground state baryon wave functions \cite{Durand:2001zz, Durand:2001sz}. 
	
	\subsection{Radiative Decay Widths}
	In this subsection, we present the predictions for M1 radiative decay widths for the charm and bottom baryons in EMS. It may be noted that currently, the experimental numbers for radiative decay widths of these baryons are not available. However, there exist a number of theoretical estimates for the same. We list our observations as follows: 
	
	\begin{enumerate}
		\item[i.] Our results for the radiative M1 decay widths of singly charmed baryons are roughly of the same order when compared with the predictions of the BM \cite{Bernotas:2013eia, Simonis:2018rld}, HBChPT \cite{Wang:2018cre}, and HCM \cite{Majethiya:2009vx}, although there are some discrepancies among the results from different models. It may be noted that to compare our results with pion mean-field approach \cite{Yang:2019tst}, we have used their results for transition moments to obtain M1 decay widths. Our results are in fairly good agreement with Yang \textit{et al.} \cite{Yang:2019tst} with few exceptions.  Interestingly, the results for $\Sigma_c^{*+}\rightarrow \Sigma_c^{+}$ decay width in various models are not only widely varying, but are also expected to have larger uncertainty arising from photon momentum. The observation of such decays can be marked as a test for various models. 
	
		\item [ii.] It is worth noting that the mass difference between $\Sigma_c^{*}$ and $\Sigma_c$, $\Xi_c^{*}$ and $\Xi_c^{'}$, $\Omega_c^{*}$ and  $\Omega_c$, $\Sigma_b^{*}$ and $\Sigma_b$, $\Xi_b^{*}$ and $\Xi_b^{'}$, $\Omega_b^{*}$ and $\Omega_b$ is less than the pion mass and because of this $\Sigma_c^{*}$,  $\Xi_c^{*}$, $\Omega_c^{*}$, $\Sigma_b^{*}$, $\Xi_b^{*}$, $\Omega_b^{*}$ should decay radiatively with small photon momenta. As already discussed, the photon momenta could be accurately given for the known experimental masses. Thus, the maximum error for the radiative decays involving singly heavy quark is up to 14\% for both, $\Sigma_c^{*+}\rightarrow \Sigma_c^{+}$, and $\Omega_b^{*-}\rightarrow \Omega_b^{-}$, which propagates mainly from larger uncertainties in photon momenta. Moreover, the absence of experimentally determined masses of heavy baryons, especially, doubly and triply heavy baryons, can be seen as a source of uncertainty in various models. Aforementioned, we have used the predictions from LQCD \cite{Brown:2014ena} as numerical values for these unobserved masses. Thus, we expect our predictions to be more reliable. 
		
		\item[iii.] Our predictions for the singly heavy bottom baryon decay widths agree qualitatively with the BM \cite{Bernotas:2013eia, Simonis:2018rld}, HBChPT \cite{Wang:2018cre}, however, are in good agreement with  pion mean-field approach \cite{Yang:2019tst}.  Exceptions may be noted in the case of LCSR \cite{Aliev:2009jt,Aliev:2014bma,Aliev:2016xvq} results for the 	$ \Sigma_b^{*}\rightarrow \Sigma_b^{}$ and $\Xi_b^{*}\rightarrow \Xi_b^{'}$ type decays for which the predictions differ by an order of magnitude. On the other hand, the decay width of $\Xi_b^{'0}\rightarrow \Xi_b^{0}$ is in excellent agreement with LCSR \cite{Aliev:2009jt, Aliev:2014bma, Aliev:2016xvq}.
		
		\item [iv.] In the case of doubly charmed M1 transition decay widths, our results are consistent with  \cite{Bernotas:2013eia, Simonis:2018rld}. For the doubly heavy baryon (involving one or more \textit{b}-quarks), we predict relatively small decay widths for $\Xi_{cb}^{'0}\rightarrow \Xi_{cb}^{0}$,  $\Xi_{cb}^{'+}\rightarrow \Xi_{cb}^{+}$, $\Xi_{cb}^{*0}\rightarrow \Xi_{cb}^{0}$, $\Xi_{cb}^{*+}\rightarrow \Xi_{cb}^{+}$  transitions. However, the results involving $\Omega_{cb}^{'(*)}$ decay channels are of the same order as the BM \cite{Bernotas:2013eia, Simonis:2018rld} results. Here again, the larger uncertainties, up to 11\%, can be seen in radiative decays involving $\Xi_{cb}^{'}$ due to larger uncertainties in photon momenta. In comparison with other models, we find that different approaches lead to different results. Some of these predictions vary by an order of magnitude, for example, the choice of the wavefunction in the approaches followed by \cite{Li:2017pxa, Lu:2017meb} yield larger decay widths as compared to the rest. Thus, experimental observation of doubly heavy baryons can reduce the theoretical ambiguities in this sector. Furthermore, it has been reported that hyperfine mixing can show a notable change in electromagnetic decay widths in light of heavy quark spin symmetry \cite{Albertus:2010hi}. Therefore, experimental measurements of electromagnetic decay widths of doubly heavy baryons could prove to be significant in the determination of hyperfine mixing in these states.
		
		\item[v.] In our approach, the radiative decay widths  of triply heavy transition for $\Omega_{ccb}^{*+}\rightarrow \Omega_{ccb}^{+}$ and $\Omega_{cbb}^{*0}\rightarrow \Omega_{cbb}^{0}$ are consistent with \cite{Simonis:2018rld}.\\
	\end{enumerate}
	
	\section{Summary and Conclusions}
		In the present work, we have primarily focused on the prediction of magnetic properties of heavy flavor baryons in the framework of EMS. The EMS takes into account the modification based on hyperfine interaction between constituent quarks via one gluon exchange inside the baryon. Another unique feature of EMS is that it does not depend on multiple parameters and, in addition, we have also incorporated symmetry breaking (through masses), which is a desirable feature for consistent predictions of baryon properties. We have calculated the magnetic and transition moments involving low-lying heavy baryons containing up to three heavy quarks, and consequently, have predicted M1 radiative decay widths for $\frac{1}{2}^{\prime} \rightarrow \frac{1}{2}$ and $\frac{3}{2} \rightarrow \frac{1}{2}^{(')}$ baryon states. Also, we have compared our results with existing predictions from other theoretical models. In order to make robust predictions, we have utilized precisely measured experimental values of baryon masses, and have used LQCD estimates in the case of unobserved baryons. Furthermore, we have improved upon the previous work by determining hyperfine interaction term ($b_{ij}$) for $s-$, $c-$, and $b-$flavors from precise experimental values of baryon masses within the same flavor sector. Following the current approach, we have accomplished two improvements. Firstly, symmetry breaking (through masses and interaction terms) is partially incorporated, and secondly, a more reliable calculation of effective masses has been achieved. In addition, we have tried to limit the uncertainties in photon momenta by mostly relying on experimental information. In the light of preceding arguments, it is therefore expected that our results would provide reasonably accurate predictions of magnetic (transition) moments and M1 radiative decay widths. We list our major findings below:
\begin{itemize}
	\item In the light of the aforementioned improvements, we have obtained transition moments and consequently, predicted the radiative decay widths of singly, doubly, and triply heavy charmed and bottom baryons. Our predictions are in good agreement with other models, with few exceptions, particularly for singly heavy baryons. The discrepancies among various models are largely due to the lack of experimental information in the heavy flavor sector.   
	\item It is interesting to note that in the case of doubly and triply heavy baryons, results from all the approaches are roughly of the same order but differ in magnitude. This indicates that the magnetic moments of heavy baryons are controlled by heavy quark magnetic moments. The smaller values of hyperfine interaction term $b_{QQ}$ for doubly and triply heavy baryons reaffirm the fact that magnetic moments of these baryons will be smaller due to smaller radii. Similar conclusions have also been reached in the BM \cite{Simonis:2018rld, Bernotas:2013eia}.
	\item The observed discrepancies between the various theoretical approaches in the predictions of magnetic and transition moments of doubly heavy baryons can mainly be attributed to the choice of wavefunctions and state mixing effects. The inclusion of diquark correlations and state mixing effects in some models lead to substantial variation in numerical values. 
	\item In singly and doubly heavy sectors, our predictions of radiative decay widths are consistent with the other theoretical approaches. Our results clearly indicate that the M1 decay widths resulting from the transition states with the mass difference less than pion mass will decay with a smaller width. Therefore, the uncertainty in the photon momenta in such decays can play a decisive role. As pointed out earlier we have taken care of such uncertainties to some extent by relying on experimental information.
\end{itemize}
	Future experimental efforts on properties of heavy flavor baryon can resolve discrepancies among different model predictions. We hope that our results prove to be useful in future experimental as well as theoretical ventures concerning heavy baryons.
\section*{Acknowledgment}
The authors gratefully acknowledge the financial support by the Department of Science and Technology (SERB:CRG/2018/002796), New Delhi.

		
\newpage

\begin{table}
	\begin{center}
		\caption{\centering Magnetic moments (in nuclear magneton, $\mu_N$) of $J^P={1/2}^+$  charm baryons.}
		\label{t1}
		\setlength{\tabcolsep}{2pt}
		\begin{tabular}{|c|c|c|c|c|c|c|c|c|} 
			\hline
			\textbf{Baryons} & \textbf{EMS} & \bf\cite{Simonis:2018rld} & \bf\cite{Yang:2018uoj} & \bf\cite{Can:2013tna} & \bf\cite{Gandhi:2018lez} & \bf\cite{Xiabng:2018qsd} & \bf\cite{Liu:2018euh} & \bf\cite{Ozdem:2018uue} \\
			\hline \hline
			$\Lambda_c^{+}$ &  $0.3801\pm0.0008$& 0.335 & - & - & 0.421& $-0.232$ & - & - \\
			\hline
			$\Sigma_c^{++}$ & $2.0932\pm0.0032$& 2.280 & $2.15\pm0.10$ & $2.027\pm0.390$ &  1.831 & 1.604 & - & - \\
			\hline
			$\Sigma_c^{+}$ &$0.4270\pm0.0017$& 0.487 &  $0.46\pm0.03$ & - &  0.380 & 0.100 & - & - \\
			\hline
			$\Sigma_c^{0}$ & $-1.2392\pm0.0017$ & $-1.310$ &  $-1.24\pm0.05$ & $-1.117\pm0.198$ & $-1.091$& $-1.403$ & - & - \\
			\hline
			$\Xi_c^{+}$ & $0.3801\pm0.0008$& 0.142 & -& - & - & 0.233 & - & - \\
			\hline
			$\Xi_c^{0}$ & $0.3801\pm0.0008$& 0.346 & - & - & - & 0.193 & - & - \\
			\hline
			$\Xi_c^{'+}$ & $0.6168\pm0.0018$ & 0.825 &  $0.60\pm0.02$ & - & 0.523 & 0.559 & - & - \\
			\hline
			$\Xi_c^{'0}$ &  $-1.0734\pm0.0013$& $-1.130$ &  $-1.05\pm0.04$ & - & $-1.012$  & $-1.077$ & - & - \\
			\hline
			$\Omega_c^{0}$ & $-0.9057\pm0.0021$& $-0.950$ &  $-0.85\pm0.05$ & $-0.639\pm0.088$ & $-1.179$& $-0.748$ & - & - \\
			\hline\hline
			$\Xi_{cc}^{++}$  &$-0.1046\pm0.0021$& $-0.110$ & - & - & - & - & - & $-0.23 \pm 0.05$ \\
			\hline
			$\Xi_{cc}^{+}$ &  $0.8148\pm0.0018$& 0.719 & -  & $0.425\pm0.029$ & - & - & $0.392\pm0.013$ & $0.43\pm0.09$ \\
			\hline
			$\Omega_{cc}^{+}$ & $0.7109\pm0.0017$& 0.645 & - & $0.413\pm0.024$ & - & - & $0.397\pm0.015$ & $0.39\pm 0.09$ \\
			\hline			\end{tabular}
	\end{center}
\end{table}


\begin{table}[ht!]
	\centering
	\captionof{table}{Magnetic moments (in $\mu_N$) of  $J^P={1/2}^+$ bottom baryons.} 
	\label{t2}
	\setlength{\tabcolsep}{2pt}
	\begin{tabular}{|c|c|c|c|c|c|c|c|} 
		\hline
		\bf{Baryons} & \textbf{EMS} & \bf\cite{Bernotas:2012nz, Franklin:1981rc} & \bf\cite{Barik:1984tq}& \bf\cite{Albertus:2006ya} & \bf\cite{Bernotas:2012nz} & \bf\cite{Simonis:2018rld} & \bf\cite{Meng:2018gan} \\
		\hline \hline
		$\Lambda_b^{0}$  & $-0.06202\pm0.00001$\footnote[1]{For very small errors the value is given up to fifth decimal place.}& $-0.060$  & -& - & $-0.066$ & $-0.060$ & - \\
		\hline
		$\Sigma_b^{+}$ &$2.1989\pm0.0021$& 2.500 & 2.575 & - & 1.622 & 2.250 & 1.590 \\
		\hline
		$\Sigma_b^{0}$  &$0.5653\pm0.0011$& 0.640 & 0.659& - & 0.422 & 0.603 & 0.390 \\
		\hline
		$\Sigma_b^{-}$  &$-1.0684\pm0.0011$& $-1.220$ & $-1.256$ & - & $-0.778$ & $-1.150$ & $-0.810$ \\
		\hline
		$\Xi_b^{0}$  &$-0.06202\pm0.00001$& $-0.110$ & -& - & $-0.100$ & $-0.106$ & $0.400$ \\
		\hline
		$\Xi_b^{-}$ &$-0.06202\pm0.00001$& $-0.050$ & -& - & $-0.063$ & $-0.056$ & $-0.730$ \\
		\hline
		$\Xi_b^{'0}$  & $0.7490\pm0.0015$  & 0.900 & 0.930& - & 0.556 & 0.782 & - \\
		\hline
		$\Xi_b^{'-}$ &$-0.9077\pm0.0012$& $-1.020$ & $-0.985$ & - & $-0.660$ & $-0.968$ & - \\
		\hline
		$\Omega_b^{-}$ &$-0.7454\pm0.0024$& $-0.790$ & $-0.714$ & - & $-0.545$ & $-0.806$ & $-0.650$ \\
		\hline\hline
		$\Xi_{cb}^{+}$ &$-0.06202\pm0.00001$ & $-0.250$ & -& $-0.475^{+0.040}_{-0.088}$ & $-0.157$ & $-0.222$ & - \\
		\hline
		$\Xi_{cb}^{0}$  & $-0.06202\pm0.00001$ & 0.130 & -& $0.518^{+0.048}_{-0.020}$ & 0.068 & 0.102 & - \\
		\hline
		$\Xi_{cb}^{'+}$  & $1.4197\pm0.0014$& 1.710 & 1.525& $1.990^{+0.270}_{-0.140}$ & 1.093 & 1.460 & - \\
		\hline
		$\Xi_{cb}^{'0}$  &$-0.2997\pm0.0008$& $-0.530$ & $-0.390$ & $-0.993^{+0.065}_{-0.137}$ & $-0.236$ & $-0.452$ & - \\
		\hline
		$\Omega_{cb}^{0}$  &$-0.06202\pm0.00001$& 0.080  & - & $0.368^{+0.010}_{-0.011}$ & 0.034 & 0.058 & - \\
		\hline
		$\Omega_{cb}^{'0}$ &$-0.1120\pm0.0012$& $-0.270$ & $-0.119$ & $-0.542^{+0.021}_{-0.024}$ & $-0.106$ & $-0.275$ & - \\
		\hline
		$\Xi_{bb}^{0}$ &$-0.6699\pm0.0006$& $-0.700$ & $-0.722$ & $-0.742^{+0.044}_{-0.091}$ & $-0.432$ & $-0.581$ & - \\
		\hline
		$\Xi_{bb}^{-}$ &$0.2108\pm0.0003$& 0.230 & 0.236 & $0.251^{+0.045}_{-0.021}$ & 0.086 & 0.171 & - \\
		\hline
		$\Omega_{bb}^{-}$  &$0.1135\pm0.0008$& 0.120  & 0.100 & $0.101^{+0.007}_{-0.007}$ & 0.043 & 0.112 & - \\
		\hline\hline
		$\Omega_{ccb}$ &$0.5261\pm0.0015$& 0.540 & 0.476& - & 0.505 & 0.455 & - \\
		\hline
		$\Omega_{cbb}$ & $-0.2096\pm0.0003$ & $-0.210$ & $-0.197$ & - & $-0.205$ & $-0.187$ & - \\
		\hline    
	\end{tabular}
\end{table}


\begin{table}[ht!]
	\centering
	\captionof{table}{ Magnetic moments (in $\mu_N$) of  $J^P={3/2}^+$ charm baryons.}
	\label{t3}
	\setlength{\tabcolsep}{.1pt}
	\begin{tabular}{|c|c|c|c|c|c|c|c|} 
		\hline 
		\textbf{Baryons}  & \textbf{EMS} & \bf\cite{Simonis:2018rld} & \bf\cite{Meng:2018gan}  & \bf\cite{Yang:2018uoj} & \bf\cite{Aliev:2008sk} & \bf\cite{Gandhi:2018lez}&  \bf\cite{Ozdem:2019zis} \\
		\hline \hline
		$\Sigma_c^{*++}$ &$3.5730\pm0.0040$& 3.980 & 2.410 &  $3.22\pm0.15$ & $4.81\pm1.22$  & 3.232 & - \\ 
		\hline
		$\Sigma_c^{*+}$  &$1.1763\pm0.0020$& 1.250 & 0.670  & $0.68\pm0.04$ & $2.00\pm0.46$ & 1.136 & - \\
		\hline
		$\Sigma_c^{*0}$  &$ -1.2198\pm0.0018$& $-1.490$ & $-1.070$  & $-1.86\pm0.07$ & $-0.81\pm0.20$ & $-1.044$ & - \\
		\hline
		$\Xi_c^{*+}$  &$1.4426\pm0.0022$& 1.470 & 0.810 & $0.90\pm0.04$ & $1.68\pm0.42$ & 1.333& -\\
		\hline
		$\Xi_c^{*0}$  &$-0.9866\pm0.0017$& $-1.200$ & $-0.900$ & $-1.
		57\pm0.06$ & $-0.68\pm0.18$ & $-0.837$ & - \\
		\hline
		$\Omega_c^{*0}$ &$ -0.7512\pm0.0029$& $-0.936$ & $-0.700$ & $-1.28\pm0.08$ & $-0.62\pm0.18$ & $ -1.129$& - \\
		\hline\hline
		$\Xi_{cc}^{*++}$ &$2.4344\pm0.0033$& 2.350 & - & - & - & - & $2.94\pm0.95$ \\
		\hline
		$\Xi_{cc}^{*+}$ &$  -0.0846\pm0.0025$& $-0.178$ & -  & - & - & - & $-0.67\pm0.11$ \\
		\hline
		$\Omega_{cc}^{*+}$ &$0.1871\pm0.0026$& 0.048 & -  & - & - & - & $-0.52\pm0.07$ \\
		\hline\hline
		$\Omega_{ccc}^{*++}$ &$1.1320\pm0.0060$& 0.989 & - & - & - & - & - \\
		\hline 
	\end{tabular}
\end{table}


\begin{table}[ht!]
	\centering
	\captionof{table}{Magnetic moments (in $\mu_N$) of $J^P={3/2}^+$ bottom baryons.} 
	\label{t4}
	\setlength{\tabcolsep}{2pt}
	\begin{tabular}{|c|c|c|c|c|c|c|c|}
		\hline 
		\textbf{Baryons}  & \textbf{EMS} & \bf\cite{Albertus:2006ya} & \bf\cite{Bernotas:2012nz} &  \bf\cite{Simonis:2018rld} & \bf\cite{Meng:2018gan} & \bf\cite{Aliev:2008sk} & \bf\cite{Ozdem:2019zis}\\
		\hline \hline
		$\Sigma_b^{*+}$ & $3.1637\pm0.0031$& - & 2.346 & 3.460 & 2.140 & $2.52\pm 0.50$ & - \\
		\hline
		$\Sigma_b^{*0}$  &$0.7444\pm0.0016$& - & 0.537 & 0.820 & 0.400 & $0.50\pm 0.15$ & - \\
		\hline
		$\Sigma_b^{*-}$  &$-1.6748\pm0.0015$& - & $-1.271$ & $-1.820$ & $-1.350$ & $-1.50\pm 0.36$ & - \\
		\hline
		$\Xi_b^{*0}$  &$1.0151\pm0.0019$& - & 0.690 & 1.030 & 0.540 & $0.50\pm 0.15$ & - \\
		\hline
		$\Xi_b^{*-}$  & $-1.4379\pm0.0015$& - & $-1.088$ & $-1.550$ & $-1.170$ & $-1.42\pm 0.35$ & -\\
		\hline
		$\Omega_b^{*-}$  &$-1.1985\pm0.0030$& - & $-0.919$ & $-1.310$ & $-0.970$ & $-1.40\pm 0.35$ & - \\
		\hline\hline
		$\Xi_{cb}^{*+}$  &$2.0131\pm0.0020$& $2.270^{+0.270}_{-0.140}$ & 1.414  & 1.880 & - & - & $2.63\pm0.82$ \\
		\hline
		$\Xi_{cb}^{*0}$  & $-0.5315\pm0.0012$& $-0.712^{+0.059}_{-0.133}$ & $-0.257$ & $-0.534$ & - & - & $-0.96\pm0.32$ \\
		\hline
		$\Omega_{cb}^{*0}$  &$-0.2552\pm0.0016$& $-0.261^{+0.015}_{-0.021}$ & $-0.111$ & $-0.329$ & - & - & $-1.11\pm0.33$  \\
		\hline
		$\Xi_{bb}^{*0}$  &$1.5897\pm0.0016$& $1.870^{+0.270}_{-0.130}$ & 0.916 & 1.400 & - & - & $2.30\pm0.55$\\
		\hline
		$\Xi_{bb}^{*-}$  &$-0.9809\pm0.0008$& $-1.110^{+0.060}_{-0.140}$ & $-0.652$ & $-0.880$ & - & - & $-1.39\pm0.32$ \\
		\hline
		$\Omega_{bb}^{*-}$ &$-0.6999\pm0.0017$& $-0.662^{+0.022}_{-0.024}$ & $-0.522$ & $-0.697$ & - & - & $-1.56\pm0.33$ \\
		\hline\hline
		$\Omega_{ccb}^{*+}$  &$0.6952\pm0.0023$& - & 0.659 & 0.594 & - & - & - \\
		\hline
		$\Omega_{cbb}^{*0}$  &$0.2558\pm0.0008$& - & 0.225 & 0.204 & - & - & - \\
		\hline
		$\Omega_{bbb}^{*-}$  &$-0.1860 \pm0.0001$& - & $-0.194$ & $-0.178$ & - & - & - \\
		\hline 
	\end{tabular}
\end{table}


\begin{table}
	\centering
	\captionof{table} {Expressions for magnetic $\mu_{\frac{1}{2}^{'+}  \to \frac{1}{2}^{+}}$ transition moments (in $\mu_N$) of charm and bottom baryons.}
	\label{t5}
	\begin{tabular}{|c|c|} \hline 
		\textbf{Transitions} & \textbf{Transition moments} \\ \hline \hline
		$\Sigma _{c}^{+}\rightarrow \Lambda _{c}^{+} $ & $\sqrt\frac{1}{3}(\mu^\xi _{d} -\mu^\xi _{u} )$\\ \hline 
		$\Xi _{c}^{'0}\rightarrow \Xi _{c}^{0} $ & $\sqrt\frac{1}{3}(\mu^\xi _{s} -\mu^\xi _{d} )$\\ \hline 
		$\Xi _{c}^{'+}\rightarrow \Xi _{c}^{+} $ & $\sqrt\frac{1}{3}(\mu^\xi _{s} -\mu^\xi _{u} )$\\ \hline\hline 
		$\Sigma _{b}^{0}\rightarrow \Lambda _{b}^{0} $ & $\sqrt\frac{1}{3}(\mu^\xi _{d} -\mu^\xi _{u} )$\\ \hline 
		$\Xi _{b}^{'-}\rightarrow \Xi _{b}^{-} $ & $\sqrt\frac{1}{3}(\mu^\xi _{s} -\mu^\xi _{d})$\\ \hline 
		$\Xi _{b}^{'0}\rightarrow \Xi _{b}^{0} $ & $\sqrt\frac{1}{3}(\mu^\xi _{s} -\mu^\xi _{u} )$\\ \hline \hline
		$\Xi _{cb}^{'0}\rightarrow \Xi _{cb}^{0} $ & $\sqrt\frac{1}{3}(\mu^\xi _{c} -\mu^\xi _{d} )$ \\ \hline 
		$\Xi _{cb}^{'+}\rightarrow \Xi _{cb}^{+} $ & $\sqrt\frac{1}{3}(\mu^\xi _{c} -\mu^\xi _{u} )$ \\ \hline 
		$\Omega _{cb}^{'0}\rightarrow \Omega _{cb}^{0} $ & $\sqrt\frac{1}{3}(\mu^\xi _{c} -\mu^\xi _{s} )$\\ \hline  
	\end{tabular}
\end{table}


\begin{table}
	\centering
	\captionof{table} {Expressions for magnetic $\mu_{\frac{3}{2}^{+}  \to \frac{1}{2}^{(')+}}$ transition moments (in $\mu_N$) of bottom baryons.}
	\label{t6}
	\begin{tabular}{|c|c|} \hline 
		\textbf{Transitions} & \textbf{ Transition moments} \\ \hline \hline
		$\Sigma _{b}^{*0}\rightarrow \Lambda _{b}^{0} $ & $\sqrt\frac{2}{3}(\mu^\xi _{u} -\mu^\xi _{d} )$\\ 
		\hline 
		$\Sigma _{b}^{*-}\rightarrow \Sigma _{b}^{-} $ & $\frac{2\sqrt{2}}{3}(\mu^\xi _{d} -\mu^\xi _{b} )$\\ 
		\hline 
		$\Sigma _{b}^{*0}\rightarrow \Sigma _{b}^{0} $  & $\frac{\sqrt{2}}{3}( \mu^\xi _{u} +\mu^\xi_{d} - 2\mu^\xi _{b})$\\ 
		\hline 
		$\Sigma _{b}^{*+}\rightarrow \Sigma _{b}^{+} $ & $\frac{2\sqrt{2}}{3}(\mu^\xi _{u} -\mu^\xi _{b} )$\\ 
		\hline 
		$\Xi _{b}^{*-}\rightarrow \Xi _{b}^{-} $ & $\sqrt\frac{2}{3}(\mu^\xi _{d} -\mu^\xi _{s} )$\\
		\hline  
		$\Xi _{b}^{*0}\rightarrow \Xi _{b}^{0} $ & $\sqrt\frac{2}{3}(\mu^\xi _{u} -\mu^\xi _{s} )$\\ 
		\hline
		$\Xi _{b}^{*-}\rightarrow \Xi _{b}^{\prime \,-} $ & $\frac{\sqrt{2}}{3}( \mu^\xi _{d} +\mu^\xi_{s}-2\mu^\xi _{b} )$\\ 
		\hline 
		
		$\Xi _{b}^{*0}\rightarrow \Xi _{b}^{\prime \,0} $ & $\frac{\sqrt{2}}{3}( \mu^\xi_{u}+\mu^\xi _{s}-2\mu^\xi _{b} )$\\ 
		\hline 
		$\Omega _{b}^{*-}\rightarrow \Omega _{b}^{-} $ & $\frac{2\sqrt{2}}{3}(\mu^\xi _{s} -\mu^\xi _{b} )$ \\ 
		\hline \hline 
		$\Xi _{cb}^{*0}\rightarrow \Xi _{cb}^{0} $ & $\sqrt\frac{2}{3}(\mu^\xi _{d} -\mu^\xi _{c} )$ \\ 
		\hline 
		$\Xi _{cb}^{*+}\rightarrow \Xi _{cb}^{+} $ & $\sqrt\frac{2}{3}(\mu^\xi _{u} -\mu^\xi _{c} )$\\ 
		\hline 
		$\Xi _{cb}^{*0}\rightarrow \Xi _{cb}^{\prime \,0} $ & $\frac{\sqrt{2}}{3}( \mu^\xi_{d}+\mu^\xi _{c} -2\mu^\xi _{b})$\\ 
		
		\hline 
		$\Xi _{cb}^{*+}\rightarrow \Xi _{cb}^{\prime\,+} $ & $\frac{\sqrt{2}}{3}( \mu^\xi_{u}+\mu^\xi _{c} -2\mu^\xi _{b} )$\\ 
		\hline
		$\Omega _{cb}^{*0}\rightarrow \Omega _{cb}^{0} $ & $\sqrt\frac{2}{3}(\mu^\xi _{s} -\mu^\xi _{c} )$\\ 
		\hline   
		$\Omega _{cb}^{*0}\rightarrow \Omega _{cb}^{\prime\,0} $ & $\frac{\sqrt{2}}{3}( \mu^\xi_{s}+\mu^\xi _{c}-2\mu^\xi _{b} )$\\ 
		\hline
		$\Xi _{bb}^{*-}\rightarrow \Xi _{bb}^{-} $ & $\frac{2\sqrt{2}}{3}(\mu^\xi _{b} -\mu^\xi _{d} )$ \\ 
		\hline 
		$\Xi _{bb}^{*0}\rightarrow \Xi _{bb}^{0} $ & $\frac{2\sqrt{2}}{3}(\mu^\xi _{b} -\mu^\xi _{u} )$\\ 
		\hline
		$\Omega _{bb}^{*-}\rightarrow \Omega _{bb}^{-} $ & $\frac{2\sqrt{2}}{3}(\mu^\xi _{b} -\mu^\xi _{s} )$\\ 
		\hline\hline
		$\Omega _{ccb}^{*+}\rightarrow \Omega _{ccb}^{+} $ & $\frac{2\sqrt{2}}{3}(\mu^\xi _{c} -\mu^\xi _{b} )$\\ 
		\hline
		$\Omega _{cbb}^{*0}\rightarrow \Omega _{cbb}^{0}$ & $\frac{2\sqrt{2}}{3}(\mu^\xi _{b} -\mu^\xi _{c} )$\\ 
		\hline
	\end{tabular}
\end{table}


\begin{table}[ht!]
	\centering
	\captionof{table}{Magnetic $\mu_{\frac{1}{2}^{'+}  \to \frac{1}{2}^{+}}$ transition moments (in $\mu_N$) of charm and bottom baryons.}
	\label{t7}
	\setlength{\tabcolsep}{2pt}
	\begin{tabular}{|c|c|c|c|c|c|c|}
		\hline
		\textbf{Transition} & \textbf{EMS} & \bf\cite{Sharma:2010vv} &\bf\cite{Bernotas:2012nz} & \bf\cite{Simonis:2018rld} & \bf\cite{Wang:2018cre}& \bf\cite{Yang:2019tst} \\
		\hline \hline
		$\Sigma_c^{+}\rightarrow \Lambda_c^{+}$ & $-1.6467\pm0.0015$ & 1.560 & 1.182 & $-1.480$ & $-1.380$ & $ 1.54\pm 0.06$ \\[2pt]
		\hline
		$\Xi_c^{\prime \,0}\rightarrow \Xi_c^{0}$ & $0.1862\pm0.0008$ & $-0.310$ & 0.013 & 0.034 & 0.220 & $ 0.21\pm 0.03 $ \\[2pt]
		\hline
		$\Xi_c^{\prime \,+}\rightarrow \Xi_c^{+}$ & $-1.4282\pm0.0012$ & 1.300 & 1.043 & $-1.330$ & 0.730 & $ -1.19\pm 0.06 $ \\[2pt]
		\hline\hline
		$\Sigma_b^{0}\rightarrow \Lambda_b^{0}$ &$-1.6305\pm0.0014$ & - & 1.052 & $-1.430$ & $-1.370$& $ -1.54\pm 0.0.6 $ \\[2pt]
		\hline
		$\Xi_b^{\prime \,-}\rightarrow \Xi_b^{-}$ & $0.1833\pm0.0008$ & - & 0.082  & 0.109 & 0.210 & $ -0.21\pm 0.03 $ \\[2pt]
		\hline
		$\Xi_b^{\prime \,0}\rightarrow \Xi_b^{0}$ & $-1.4150\pm0.0011$ & - & 0.917 & $-1.300$ & $-0.750$ & $1.19\pm 0.06 $ \\[2pt]
		\hline\hline
		$\Xi_{cb}^{\prime \,0}\rightarrow \Xi_{cb}^{0}$ &  $0.7295\pm0.0007$  & - & 0.508  & 0.598 & - & - \\[2pt]
		\hline
		$\Xi_{cb}^{\prime \,+}\rightarrow \Xi_{cb}^{+}$ & $-0.7983\pm0.0012$ & - & 0.277 & $-0.531$ &- & - \\[2pt]
		\hline
		$\Omega_{cb}^{\prime \,0}\rightarrow \Omega_{cb}^{0}$ & $0.5603\pm0.0007$ & - & 0.443 & 0.508 & - & - \\[2pt]
		\hline 
	\end{tabular}
\end{table}


\begin{table}[ht!]
	\centering
	\captionof{table}{ Magnetic $\mu_{\frac{3}{2}^{+}  \to \frac{1}{2}^{(')+}}$ transition moments (in $\mu_N$) of charm baryons.}
	\label{t8}
	\setlength{\tabcolsep}{1pt}
	\begin{tabular}{|c|c|c|c|c|c|c|c|c|} 
		\hline
		\textbf{Transition} & \textbf{EMS} & \bf\cite{Sharma:2010vv} & \bf\cite{Simonis:2018rld} & \bf\cite{Wang:2018cre} & \bf\cite{Yang:2019tst} & \bf\cite{Aliev:2009jt}\footnote[2]{T. M. Aliev, K. Azizi and A. Ozpineci have given their results in natural magneton ($e\hbar/ 2cM_{\mathcal{B}}$), however, to convert to nuclear magneton we multiply the entire magnetic moments with $2m_{N}/ (M_{\mathcal{B}_{{3/2}^{+}}} + M_{\mathcal{B}_{{1/2}^{+}}}).$} & \bf\cite{Gandhi:2018lez}& \bf\cite{Li:2017pxa} \\
		\hline \hline
		$\Sigma_c^{*+} \rightarrow\Lambda_c^{+}$ & $2.2802\pm0.0020$ & 2.400 & 2.070 & 2.000 & $-2.18\pm0.08$ & $1.48\pm0.55$ & 1.758 & - \\
		\hline
		$\Sigma_c^{*++} \rightarrow\Sigma_c^{++}$ & $1.1786\pm0.0015$ & $-1.370$ &  1.340 & 1.070 & $1.52\pm0.07$ & $1.06\pm0.38$ & 0.988 & - \\
		\hline
		$\Sigma_c^{*+} \rightarrow\Sigma_c^{+}$ & $0.0250\pm0.0009$ & $-0.003$ & 0.102 & 0.190& $0.33\pm0.02$ & $0.45\pm0.11$ & 0.009& - \\
		\hline
		$\Sigma_c^{*0} \rightarrow\Sigma_c^{0}$  & $-1.1286\pm0.0009$ & 1.480 & $-1.140$ & $-0.690$ & $-0.87\pm0.03$ & $0.19\pm0.08$ & 1.013 & - \\
		\hline
		$\Xi_c^{*+} \rightarrow\Xi_c^{+}$ & $1.9797\pm0.0015$ & 2.080 & 1.860 & 1.050 & $1.69\pm0.08$ & $1.47\pm0.66$ & 0.985 & - \\
		\hline
		$\Xi_c^{*0} \rightarrow\Xi_c^{0}$  & $-0.2550\pm0.0010$ & $-0.500$ & $-0.249$ & $-0.310$ & $-0.29\pm0.04$ & $ 0.16\pm0.075$ &0.253 & - \\
		\hline
		
		$\Xi_c^{*+} \rightarrow\Xi_c^{'+}$  & $0.1548\pm0.0010$ & $-0.230$ & 0.066 & 0.230 &  $0.43\pm0.02$ & - & - & - \\
		\hline
		$\Xi_c^{*0} \rightarrow\Xi_c^{'0}$  & $-1.0151\pm0.0008$ & 1.240 & $-0.994$ & $-0.590$ &  $-0.74\pm0.03$ & - &  - & - \\
		\hline  
		$\Omega_c^{*0} \rightarrow\Omega_c^{0}$  & $-0.9004\pm0.0011$ & 0.960 & $-0.892$ & $-0.490$ & $-0.60\pm0.04$ & - & 0.872 & - \\
		\hline\hline
		$\Xi_{cc}^{*++} \rightarrow\Xi_{cc}^{++}$  & $-1.2991\pm0.0024$ & 1.330 & $-1.210$ & - & - & - & - & $-2.350 $ \\
		\hline
		$\Xi_{cc}^{*+} \rightarrow\Xi_{cc}^{+}$  & $1.1861\pm0.0014$ & $-1.410$ & 1.070 & - & - & - & - & 1.550 \\
		\hline
		$\Omega_{cc}^{*+} \rightarrow\Omega_{cc}^{+}$  & $0.9111\pm0.0013$ & $-0.890$ & 0.869 & - & - & - & - & 1.540 \\
		\hline  
	\end{tabular}
\end{table}


\begin{table}[tbp] \centering
	\caption{\centering Magnetic $\mu_{\frac{3}{2}^{+}  \to \frac{1}{2}^{(')+}}$ transition moments (in $\mu_N$) of bottom baryons.}
	\label{t9}
	\setlength{\tabcolsep}{2pt}
	\begin{tabular}{|c|c|c|c|c|}
		\hline
		\bf{Transition} & \textbf{EMS} & \bf\cite{Simonis:2018rld} & \bf\cite{Wang:2018cre} & \bf\cite{Yang:2019tst} \\ 
		\hline
		\hline
		$\Sigma _{b}^{*0}\rightarrow \Lambda _{b}^{0} $ & $2.2911\pm0.0019$ & 2.020 & 1.960 & $-2.18\pm0.08$ \\
		\hline
		$\Sigma _{b}^{*-}\rightarrow \Sigma _{b}^{-} $ & $-0.7067\pm0.0005$ & $-0.760$ & $-0.580$ & $0.87\pm0.03$ \\
		\hline
		$\Sigma _{b}^{*0}\rightarrow \Sigma _{b}^{0} $ & $0.4411\pm0.0006$ & 0.464  & 0.300 & $-0.33\pm0.02$ \\
		\hline
		$\Sigma _{b}^{*+}\rightarrow \Sigma _{b}^{+} $ & $1.5889\pm0.0010$ & 1.690 & 1.170 & $-1.52\pm0.07$ \\
		\hline 
		$\Xi _{b}^{*-}\rightarrow \Xi _{b}^{-} $ & $-0.2570\pm0.0010$ & $-0.182$ & $-0.300$ & $-0.29\pm0.04$ \\
		\hline
		$\Xi _{b}^{*0}\rightarrow \Xi _{b}^{0} $ & $1.9887\pm0.0015$ & 1.830 & 1.060 & $1.69\pm0.08$ \\
		\hline
		$\Xi _{b}^{*-}\rightarrow \Xi _{b}^{\prime \,-} $ &$-0.5940\pm0.0005$ & $-0.623$ & $-0.490$ & $0.74\pm0.03$ \\
		\hline
		
		$\Xi _{b}^{*0}\rightarrow \Xi _{b}^{\prime \,0} $ &$0.5698\pm0.0007$ & 0.521 & 0.330 & $-0.43\pm0.02$ \\
		\hline
		$\Omega _{b}^{*-}\rightarrow \Omega _{b}^{-} $ & $-0.4802\pm0.0011$ & $-0.523$ & $-0.380$ & $0.60 \pm 0.04$\\
		\hline\hline
		$\Xi _{cb}^{*0}\rightarrow \Xi _{cb}^{0} $ & $-1.0267\pm0.0009$ & $-0.919$ & - & - \\
		\hline
		$\Xi _{cb}^{*+}\rightarrow \Xi _{cb}^{+} $ & $1.1195\pm0.0017$ & 1.120  & - & - \\
		\hline
		$\Xi _{cb}^{*0}\rightarrow \Xi _{cb}^{\prime \,0} $ & $-0.1654\pm0.0004$ & $-0.042$ & - & -  \\
		\hline
		
		$\Xi _{cb}^{*+}\rightarrow \Xi _{cb}^{\prime\,+} $ &$1.0422\pm0.0007$ & 0.814 & - & -  \\
		\hline
		$\Omega _{cb}^{*0}\rightarrow \Omega _{cb}^{0} $ & $-0.7896\pm0.0010$ & $-0.748$ & - & - \\
		\hline
		$\Omega _{cb}^{*0}\rightarrow \Omega _{cb}^{\prime\,0} $ & $-0.0340\pm0.0006$ & 0.017 & - & - \\
		\hline
		$\Xi _{bb}^{*-}\rightarrow \Xi _{bb}^{-} $ &  $0.7605\pm0.0006$ & 0.643  & - & - \\
		\hline
		$\Xi _{bb}^{*0}\rightarrow \Xi _{bb}^{0} $ & $-1.6965\pm0.0011$ & $-1.450$ & - & - \\
		\hline
		$\Omega _{bb}^{*-}\rightarrow \Omega _{bb}^{-} $ &$0.4906\pm0.0014$ &	0.478 & - & -  \\
		\hline
		\hline
		$\Omega _{ccb}^{*+}\rightarrow \Omega _{ccb}^{+} $ &$0.4157\pm0.0008$ & 0.362 & - & - \\
		\hline
		$\Omega _{cbb}^{*0}\rightarrow \Omega _{cbb}^{0}$ & $-0.4170\pm0.0005$ & $-0.352$ & - & - \\ 
		\hline 
	\end{tabular}
\end{table}


\begin{table}[!ht]
	\centering
	\captionof{table}{Momenta of emitted photon, $\omega$, for charm and bottom baryons.}
	\label{t10}
	\setlength{\tabcolsep}{2pt}
	\begin{tabular}{|c|c|c|c|c||c|}
		\hline 
		\textbf{Transition} &  \textbf{$\omega$ (in MeV)} & \bf\cite{Simonis:2018rld} & \textbf{Transition} & \textbf{$\omega$ (in MeV)} & \bf\cite{Simonis:2018rld} \\
		\hline
		\hline
		$\Sigma_{c}^{+}\rightarrow\Lambda_{c}^{+}$  &$160.80\pm0.40$ & 160 & $\Sigma_{b}^{0}\rightarrow\Lambda_{b}^{0}$  &$190.30\pm0.24$ & 190   \\
		\hline
		$\Xi_{c}^{\prime\,0}\rightarrow\Xi_{c}^{0}$ &$106.00\pm0.50$ & 105 &
		$\Xi_{b}^{\prime\,0}\rightarrow\Xi_{b}^{0}$ &$141.40\pm0.50$& 135 \\
		\hline
		$\Xi_{c}^{\prime\,+}\rightarrow\Xi_{c}^{+}$ &$108.10\pm0.50$ & 106 &  
		$\Xi_{b}^{\prime\,-}\rightarrow\Xi_{b}^{-}$ &$136.40\pm0.60$& 138   \\
		\hline
		\hline
		$\Sigma_{c}^{*+}\rightarrow\Lambda_{c}^{+}$ & $220.40\pm2.10$  & 220 &   $\Sigma_{b}^{*0}\rightarrow\Lambda_{b}^{0}$ &$209.00\pm0.25$ & 210  \\
		\hline
		$\Sigma_{c}^{*0}\rightarrow\Sigma_{c}^{0}$  & $63.90\pm0.24$  & 64  & 
		$\Sigma_{b}^{*0}\rightarrow\Sigma_{b}^{0}$  & $19.40\pm0.27$ & 20 \\
		\hline
		$\Sigma_{c}^{*+}\rightarrow\Sigma_{c}^{+}$  & $63.80\pm2.30$  & 64  &    
		$\Sigma_{b}^{*-}\rightarrow\Sigma_{b}^{-}$  & $19.10\pm0.40$ & 20  \\
		\hline
		
		$\Sigma_{c}^{*++}\rightarrow\Sigma_{c}^{++}$& $63.62\pm0.25$ & 64 & 
		$\Sigma_{b}^{*+}\rightarrow\Sigma_{b}^{+}$  & $19.70\pm0.40$  & 20 \\
		\hline
		$\Xi_{c}^{*0}\rightarrow\Xi_{c}^{0}$ & $169.90\pm0.35$ & 169 & 
		$\Xi_{b}^{*0}\rightarrow\Xi_{b}^{0}$ & $158.20\pm0.80$  & 155 \\
		\hline
		$\Xi_{c}^{*+}\rightarrow\Xi_{c}^{+}$ & $171.40\pm0.35$  & 172 & 
		$\Xi_{b}^{*-}\rightarrow\Xi_{b}^{-}$ & $156.20\pm0.60$ & 158  \\
		\hline
		$\Xi_{c}^{*0}\rightarrow\Xi_{c}^{\prime\,0}$ & $66.60\pm0.50$ & 67 & 
		$\Xi_{b}^{*0}\rightarrow\Xi_{b}^{\prime\,0}$ & $17.30\pm0.60$ & 20   \\
		\hline
		
		$\Xi_{c}^{*+}\rightarrow\Xi_{c}^{\prime\,+}$ &$66.10\pm0.60$ & 69 &  
		$\Xi_{b}^{*-}\rightarrow\Xi_{b}^{\prime\,-}$ & $20.28\pm0.14$ & 20 \\
		\hline
		$\Omega_{c}^{*0}\rightarrow\Omega_{c}^{0}$  & $69.80\pm2.60$ & 70 &   $\Omega_{b}^{*-}\rightarrow\Omega_{b}^{-}$  &$38.80\pm1.80$ & 20  \\
		\hline
		
	\end{tabular}
\end{table}


\begin{table}[ht!]
	\begin{center}
		\caption{Momenta of emitted photon, $\omega$, for doubly and triply heavy baryons.}
		\label{t11}
		\setlength{\tabcolsep}{2pt}
		\begin{tabular}{|c|c|}
			\hline
			\textbf{Transition} & \textbf{$\omega$ (in MeV)}\footnote[3]{All the input masses are taken from LQCD \cite{Brown:2014ena}, except for $\Xi_{cc}^{++}$. } \\
			\hline
			\hline
			$\Xi_{cb}^{'0}\rightarrow \Xi_{cb}^{0}$ &$16.00\pm0.60$\\
			\hline
			$\Xi_{cb}^{'+}\rightarrow \Xi_{cb}^{+}$ & $16.00\pm0.60$ \\
			\hline
			$\Omega_{cb}^{'0}\rightarrow \Omega_{cb}^{0}$ & $33.90\pm0.50$ \\
			\hline
			\hline
			$\Xi_{cc}^{*++}\rightarrow \Xi_{cc}^{++}$ & $69.70\pm0.50$ \\
			\hline
			$\Xi_{cc}^{*+}\rightarrow \Xi_{cc}^{+}$  & $69.70\pm0.50$ \\
			\hline
			$\Omega_{cc}^{*+}\rightarrow \Omega_{cc}^{+}$  & $83.10\pm0.40$ \\
			\hline
			$\Xi_{cb}^{*0}\rightarrow \Xi_{cb}^{0}$ & $41.90\pm0.60$ \\
			\hline
			$\Xi_{cb}^{*+}\rightarrow \Xi_{cb}^{+}$ & $41.90\pm0.60$ \\
			\hline 	
			$\Xi_{cb}^{*0}\rightarrow \Xi_{cb}^{'0}$ &$26.00\pm0.60$\\
			\hline 
			$\Xi_{cb}^{*+}\rightarrow \Xi_{cb}^{'+}$ &$26.00\pm0.60$\\
			\hline
			$\Omega_{cb}^{*0}\rightarrow \Omega_{cb}^{0}$ & $60.70\pm0.50$\\
			\hline
			$\Omega_{cb}^{*0}\rightarrow \Omega_{cb}^{'0}$ &$26.90\pm0.50$\\
			\hline
			$\Xi_{bb}^{*-}\rightarrow \Xi_{bb}^{-}$ &$34.90\pm0.50$\\
			\hline
			$\Xi_{bb}^{*0}\rightarrow \Xi_{bb}^{0}$ & $34.90\pm0.50$\\
			\hline
			$\Omega_{bb}^{*-}\rightarrow \Omega_{bb}^{-}$ &$34.90\pm0.50$\\
			\hline
			\hline
			$\Omega_{ccb}^{*+}\rightarrow \Omega_{ccb}^{+}$ & $29.94\pm0.31$\\ 
			\hline
			$\Omega_{cbb}^{*0}\rightarrow \Omega_{cbb}^{0}$ &$33.95\pm0.31$ \\
			\hline
			
		\end{tabular}	
	\end{center}	
\end{table}


\begin{table}[!ht]
	\centering
	\captionof{table}{The radiative M1 decay widths (in KeV) of charm baryons.}
	\label{t12}
	\setlength{\tabcolsep}{2pt}
	\begin{tabular}{|c|c|c|c|c|c|c|c|c|c|} 
		\hline 
		\textbf{Transition} & \textbf{EMS} & \bf\cite{Bernotas:2013eia} & \bf\cite{Simonis:2018rld} & \bf\cite{Wang:2018cre} & \bf\cite{Yang:2019tst}\footnote[4]{The values given in the column are calculated from the transition moments given by G.~S.~Yang and H.~C.~Kim in their results.} & \bf\cite{Aliev:2009jt, Aliev:2014bma, Aliev:2016xvq} & \bf\cite{Wang:2017kfr} & \bf\cite{Majethiya:2009vx} & \bf\cite{Gandhi:2018lez} \\ \hline
		\hline
		$ \Sigma_c^{+}\rightarrow \Lambda_c^{+}\gamma$  & $93.5\pm0.7$ & 46.10 & 74.10 & 65.60 & 81.05 & $50.0\pm17.0$ & 80.60 & 97.98 & 66.66\\
		\hline
		$	\Xi_c^{'0}\rightarrow \Xi_c^{0}\gamma$  & $0.342\pm0.006$& 0.002 & 0.185 & 0.460 & 0.432 & $0.27\pm0.06$ & 0000 & - & - \\
		\hline
		$   \Xi_c^{'+}\rightarrow \Xi_c^{+}\gamma$ & $21.38\pm0.31$& 10.20 & 18.60 & 5.430 & 14.78 & $8.5\pm2.5$ & 42.30 & - & -\\
		\hline
		\hline
		$  \Sigma_c^{*+}\rightarrow \Lambda_c^{+}\gamma$ &$231\pm7$& 126.0 & 190.0 & 161.6 & 211.1 & $130.0\pm 45.0$ & 373.0 & 244.4 & 135.3\\
		\hline
		$ \Sigma_c^{*++}\rightarrow \Sigma_c^{++}\gamma$  & $1.483\pm0.018$& 0.826 & 1.960 & 1.200 & 2.468 & $2.65\pm1.20$ & 3.940 & 1.980 & 2.060 \\
		\hline
		$	\Sigma_c^{*0}\rightarrow \Sigma_c^{0}\gamma$  & $1.378\pm0.016$ & 1.080 & 1.410 & 0.490 & 0.818 & $0.08\pm0.03$ & 3.430& 1.440 & 2.162\\
		\hline
		$   \Sigma_c^{*+}\rightarrow \Sigma_c^{+}\gamma$  & $(6.7 \pm 0.9)\t 10^{-4}$ & 0.004 & 0.011 & 0.040 & 0.174 & $0.40\pm0.16$ & 0.040  & 0.011 & $4\times10^{-5}$\\
		\hline
		
		$	\Xi_c^{*0}\rightarrow \Xi_c^{0}\gamma$ & $1.322\pm0.014$ & 0.908 & 0.745 & 1.840 & 1.707 & $0.66\pm0.32$ & 0000& 1.150 & 0.811\\
		\hline
		$   \Xi_c^{*+}\rightarrow \Xi_c^{+}\gamma$ & $81.9\pm0.5$ & 44.30 & 81.60 & 21.60 & 59.93 & $52\pm25$ & 139.0 & 99.94 & 15.69\\
		\hline
		$	\Xi_c^{*0}\rightarrow \Xi_c^{'0}\gamma$  & $1.262\pm0.031$ & 1.030 & 1.330 & 0.420 &  0.663 & 2.142 & 3.030 & - & - \\
		\hline
		$   \Xi_c^{*+}\rightarrow \Xi_c^{'+}\gamma$  & $0.029\pm0.001$ & 0.011 & 0.063 & 0.070 & 0.218 & 0.274 & 0.004 & - & -\\
		\hline
		$   \Omega_c^{*0}\rightarrow \Omega_c^{0}\gamma$  &$1.14\pm0.13$ & 1.070 & 1.130 & 0.320 & 0.508 & 0.932 & 0.890 & 0.820 & 0.464\\ 
		\hline
		\hline
		$   \Xi_{cc}^{*++}\rightarrow \Xi_{cc}^{++}\gamma$ & $2.37\pm0.05$ & 1.430 & 2.790 & - & - & - & - & - & -\\
		\hline
		$  \Xi_{cc}^{*+}\rightarrow \Xi_{cc}^{+}\gamma$  &$1.98\pm0.04$ & 2.080 & 2.170 & - & - & - & - & - & -\\
		\hline
		$  \Omega_{cc}^{*+}\rightarrow \Omega_{cc}^{+}\gamma$ & $1.973\pm0.029$ & 0.949 & 1.600 & - & - & - & - & - & - \\       
		\hline  
	\end{tabular}
\end{table}


\begin{table}[!ht]
	\centering
	\captionof{table}{The radiative M1 decay widths (in KeV) of bottom baryons.}
	\label{t13}
	\setlength{\tabcolsep}{2pt}
	\begin{tabular}{|c|c|c|c|c|c|c|c|}
		\hline 
		\textbf{Transition}  & \textbf{EMS} & \bf\cite{Bernotas:2013eia} & \bf\cite{Simonis:2018rld} & \bf\cite{Wang:2018cre} & \bf\cite{Yang:2019tst}\footnote[5]{The values given in the column are calculated from the transition moments given by G.~S.~Yang and H.~C.~Kim in their results.} & \bf\cite{Aliev:2009jt, Aliev:2014bma, Aliev:2016xvq} & \bf\cite{Wang:2017kfr} \\	\hline
		\hline
		$	\Sigma_b^{0}\rightarrow \Lambda_b^{0}\gamma$  &$151.9\pm0.6$ & 58.90 & 116.0 & 108.0 & 138.6 & $152.0\pm60.0$ & 130.0 \\
		\hline
		$	\Xi_b^{'-}\rightarrow \Xi_b^{-}\gamma$ &$0.707\pm0.011$ & 0.118 & 0.357 & 1.000 & 0.912 & $3.3\pm1.3$ & 0000  \\
		\hline
		$	\Xi_b^{'0}\rightarrow \Xi_b^{0}\gamma$ & $46.9\pm0.5$& 14.70 & 36.40 & 13.00 & 32.96 & $47.0\pm21.0$ & 84.60  \\
		\hline
		\hline
		$	\Sigma_b^{*0}\rightarrow \Lambda_b^{0}\gamma$  & $198.8\pm0.8$ & 81.10 & 158.0 & 142.1 & 180.0 & $114.0\pm45.0$ & 335.0 \\
		\hline
		$ \Sigma_b^{*-}\rightarrow \Sigma_b^{-}\gamma$  & $0.0144\pm0.0009$& 0.010 & 0.019 & 0.010 & 0.021 & $0.11\pm0.06$ & 0.060 \\
		\hline
		$	\Sigma_b^{*0}\rightarrow \Sigma_b^{0}\gamma$ &$0.0059\pm0.0002$ & 0.005 & 0.008 & 0.003 & 0.002 & $0.028\pm0.016$ & 0.020 \\
		\hline
		$	\Sigma_b^{*+}\rightarrow \Sigma_b^{+}\gamma$ &$0.080\pm0.004$ & 0.054 & 0.110 & 0.050 & 0.073 & $0.46\pm0.22$ & 0.250 \\
		\hline
		$	\Xi_b^{*-}\rightarrow \Xi_b^{-}\gamma$ &$1.044\pm0.015$ & 0.278 & 0.536 & 1.400 & 1.332 & $1.50\pm 0.75$ & 0000 \\
		\hline
		$	\Xi_b^{*0}\rightarrow \Xi_b^{0}\gamma$ &$65.0\pm0.9$& 24.70 & 55.30 & 17.20 & 46.92 & $135.0\pm65.0$ & 104.0 \\
		\hline
		$	\Xi_b^{*-}\rightarrow \Xi_b^{'-}\gamma$  &$0.0122\pm0.0002$& 0.005 & 0.014 & 0.008 & 0.019 & 0.303 & 15.00 \\
		\hline
		$	\Xi_b^{*0}\rightarrow \Xi_b^{'0}\gamma$ &$0.0069\pm0.0007$ & 0.004 & 0.010 & 0.002 & 0.004 & 0.131 & 5.190  \\
		\hline
		$	\Omega_b^{*-}\rightarrow \Omega_b^{-}\gamma$ &$0.056\pm0.008$& 0.006 & 0.009 & 0.031 & 0.088 & 0.092 & 0.100  \\
		\hline 
	\end{tabular}
\end{table}


\begin{table}[!ht]
	\centering
	\captionof{table}{The radiative M1 decay widths (in KeV) of bottom baryons.}
	\label{t14}
	\setlength{\tabcolsep}{2pt}
	\begin{tabular}{|c|c|c|c|c|c|} 
		\hline 
		\textbf{Transition} & \textbf{EMS} &\bf\cite{Bernotas:2013eia} & \bf \cite{Simonis:2018rld} & \bf\cite{Li:2017pxa} & \bf\cite{Lu:2017meb} \\  
		\hline
		\hline
		$	\Xi_{cb}^{'0}\rightarrow \Xi_{cb}^{0}\gamma$ & $0.0180\pm0.0021$& 0.125 & 0.204 & - & - \\
		\hline
		$	\Xi_{cb}^{'+}\rightarrow \Xi_{cb}^{+}\gamma$  & $0.0216\pm0.0025$& 0.037 & 0.161 & - & - \\
		\hline
		$	\Omega_{cb}^{'0}\rightarrow \Omega_{cb}^{0}\gamma$  & $0.102\pm0.004$& 0.053 & 0.170 & - & - \\ 
		\hline
		\hline
		$\Xi_{cb}^{*0}\rightarrow \Xi_{cb}^{0}\gamma$ & $0.321\pm0.014$ & 0.612 &  0.876 & 0.520 & - \\
		\hline
		$	\Xi_{cb}^{*+}\rightarrow \Xi_{cb}^{+}\gamma$  & $0.381\pm0.017$ & 0.533 & 1.310 & 0.520 & - \\
		\hline
		$	\Xi_{cb}^{*0}\rightarrow \Xi_{cb}^{\prime\,0}\gamma$  &$0.0020\pm0.0002$& $3\times 10^{-4}$ & $7.6\times10^{-5}$ & 7.190 & - \\
		\hline
		$	\Xi_{cb}^{*+}\rightarrow \Xi_{cb}^{\prime\,+}\gamma$  & $0.079\pm0.006$& 0.031 & 0.029 & 26.20 & - \\
		\hline
		$	\Omega_{cb}^{*0}\rightarrow \Omega_{cb}^{0}\gamma$  & $0.579\pm0.014$& 0.239 & 0.637 & 0.520 & - \\ 
		\hline
		$	\Omega_{cb}^{*0}\rightarrow \Omega_{cb}^{\prime\,0}\gamma$  & $(9.4\pm 0.6)\times10^{-5}$ & $5\times10^{-4}$ & $1.3\times10^{-5}$ & 7.080 & - \\ 
		\hline
		$	\Xi_{bb}^{*-}\rightarrow \Xi_{bb}^{-}\gamma$  & $0.102\pm0.005$& 0.022 & 0.027 & 5.170 & 0.210 \\
		\hline
		$	\Xi_{bb}^{*0}\rightarrow \Xi_{bb}^{0}\gamma$ & $0.509\pm0.023$& 0.126 & 0.137 & 31.10 & 0.980 \\
		\hline
		$	\Omega_{bb}^{*-}\rightarrow \Omega_{bb}^{-}\gamma$  &$0.0426\pm0.0018$& 0.011 & 0.015 & 5.080 & 0.040 \\ 
		\hline
		\hline
		$	\Omega_{ccb}^{*+}\rightarrow \Omega_{ccb}^{+}\gamma$  & $0.0192\pm0.0006$& 0.004 & 0.010 & - & - \\ 
		\hline
		$	\Omega_{cbb}^{*0}\rightarrow \Omega_{cbb}^{0}\gamma$  & $0.0282\pm0.0008$& 0.005 & 0.013 & - & - \\ 
		\hline 
	\end{tabular}
\end{table}



\newpage	

\begin{thebibliography}{}
\section*{References}


\bibitem{Aaij:2017nav}
R.~Aaij \textit{et al.} [LHCb],
Phys. Rev. Lett. \textbf{118}, no.18, 182001 (2017)
doi:10.1103/PhysRevLett.118.182001
[arXiv:1703.04639 [hep-ex]].

\bibitem{Aaij:2017ueg}
		R.~Aaij \textit{et al.} [LHCb],
		Phys. Rev. Lett. \textbf{119}, no.11, 112001 (2017)
		doi:10.1103/PhysRevLett.119.112001
		[arXiv:1707.01621 [hep-ex]].

		\bibitem{Aaij:2018wzf}
		R.~Aaij \textit{et al.} [LHCb],
		Phys. Rev. Lett. \textbf{121}, no.5, 052002 (2018)
		doi:10.1103/PhysRevLett.121.052002
		[arXiv:1806.02744 [hep-ex]].

\bibitem{Aaij:2018yqz}
R.~Aaij \textit{et al.} [LHCb],
Phys. Rev. Lett. \textbf{121}, no.7, 072002 (2018)
doi:10.1103/PhysRevLett.121.072002
[arXiv:1805.09418 [hep-ex]].


		\bibitem{Aaij:2018tnn}
		R.~Aaij \textit{et al.} [LHCb],
		Phys. Rev. Lett. \textbf{122}, no.1, 012001 (2019)
		doi:10.1103/PhysRevLett.122.012001
		[arXiv:1809.07752 [hep-ex]].
	    
	    
\bibitem{Aaij:2019amv}
R.~Aaij \textit{et al.} [LHCb],
Phys. Rev. Lett. \textbf{123}, no.15, 152001 (2019)
doi:10.1103/PhysRevLett.123.152001
[arXiv:1907.13598 [hep-ex]].
		
		        
\bibitem{Sirunyan:2020gtz}
A.~M.~Sirunyan \textit{et al.} [CMS],
Phys. Lett. B \textbf{803}, 135345 (2020)
doi:10.1016/j.physletb.2020.135345
[arXiv:2001.06533 [hep-ex]].
        
        
\bibitem{Aaij:2020rkw}
R.~Aaij \textit{et al.} [LHCb],
JHEP \textbf{06}, 136 (2020)
doi:10.1007/JHEP06(2020)136
[arXiv:2002.05112 [hep-ex]].
  	
\bibitem{Aaij:2020cex}
        R.~Aaij \textit{et al.} [LHCb],
        Phys. Rev. Lett. \textbf{124}, no.8, 082002 (2020)
        doi:10.1103/PhysRevLett.124.082002
        [arXiv:2001.00851 [hep-ex]].
 
 
\bibitem{Sirunyan:2021vxz}
A.~M.~Sirunyan \textit{et al.} [CMS],
[arXiv:2102.04524 [hep-ex]].


\bibitem{Aaij:2016jnn}
		R.~Aaij \textit{et al.} [LHCb],
		JHEP \textbf{05}, 161 (2016)
		doi:10.1007/JHEP05(2016)161
		[arXiv:1604.03896 [hep-ex]].
		
\bibitem{Aaij:2017awb}
		R.~Aaij \textit{et al.} [LHCb],
		Phys. Rev. Lett. \textbf{119}, no.6, 062001 (2017)
		doi:10.1103/PhysRevLett.119.062001
		[arXiv:1704.07900 [hep-ex]].
		
		
\bibitem{Aaij:2019ezy}
		R.~Aaij \textit{et al.} [LHCb],
		Phys. Rev. D \textbf{99}, no.5, 052006 (2019)
		doi:10.1103/PhysRevD.99.052006
		[arXiv:1901.07075 [hep-ex]].
		
\bibitem{Zyla:2020zbs}
P.~A.~Zyla \textit{et al.} [Particle Data Group],
PTEP \textbf{2020}, no.8, 083C01 (2020)
doi:10.1093/ptep/ptaa104
 
\bibitem{Chen:2016spr}
H.~X.~Chen, W.~Chen, X.~Liu, Y.~R.~Liu and S.~L.~Zhu,
Rept. Prog. Phys. \textbf{80}, no.7, 076201 (2017)
doi:10.1088/1361-6633/aa6420
[arXiv:1609.08928 [hep-ph]].

  

\bibitem{Aaij:2020vid}
R.~Aaij \textit{et al.} [LHCb],
JHEP \textbf{11}, 095 (2020)
doi:10.1007/JHEP11(2020)095
[arXiv:2009.02481 [hep-ex]].
  
\bibitem{DeRujula:1975qlm}
		A.~De Rujula, H.~Georgi and S.~L.~Glashow,
		Phys. Rev. D \textbf{12}, 147-162 (1975)
		doi:10.1103/PhysRevD.12.147
		
\bibitem{Brambilla:2014jmp}
N.~Brambilla, S.~Eidelman, P.~Foka, S.~Gardner, A.~S.~Kronfeld, M.~G.~Alford, R.~Alkofer, M.~Butenschoen, T.~D.~Cohen and J.~Erdmenger, \textit{et al.}
Eur. Phys. J. C \textbf{74}, no.10, 2981 (2014)
doi:10.1140/epjc/s10052-014-2981-5
[arXiv:1404.3723 [hep-ph]].

\bibitem{Aubert:2006je}
B.~Aubert \textit{et al.} [BaBar],
Phys. Rev. Lett. \textbf{97}, 232001 (2006)
doi:10.1103/PhysRevLett.97.232001
[arXiv:hep-ex/0608055 [hep-ex]].

\bibitem{Solovieva:2008fw}
E.~Solovieva, R.~Chistov, I.~Adachi, H.~Aihara, K.~Arinstein, T.~Aushev, A.~M.~Bakich, V.~Balagura, U.~Bitenc and A.~Bondar, \textit{et al.}
Phys. Lett. B \textbf{672}, 1-5 (2009)
doi:10.1016/j.physletb.2008.12.062
[arXiv:0808.3677 [hep-ex]].

\bibitem{Jessop:1998wt}
C.~P.~Jessop \textit{et al.} [CLEO],
Phys. Rev. Lett. \textbf{82}, 492-496 (1999)
doi:10.1103/PhysRevLett.82.492
[arXiv:hep-ex/9810036 [hep-ex]].

\bibitem{Aubert:2006rv}
B.~Aubert \textit{et al.} [BaBar],
[arXiv:hep-ex/0607086 [hep-ex]].

\bibitem{Yelton:2016fqw}
J.~Yelton \textit{et al.} [Belle],
Phys. Rev. D \textbf{94}, no.5, 052011 (2016)
doi:10.1103/PhysRevD.94.052011
[arXiv:1607.07123 [hep-ex]].

\bibitem{Ablikim:2019hff}
M.~Ablikim \textit{et al.} [BESIII],
Chin. Phys. C \textbf{44}, no.4, 040001 (2020)
doi:10.1088/1674-1137/44/4/040001
[arXiv:1912.05983 [hep-ex]].

\bibitem{Yuan:2019zfo}
C.~Z.~Yuan and S.~L.~Olsen,
Nature Rev. Phys. \textbf{1}, no.8, 480-494 (2019)
doi:10.1038/s42254-019-0082-y
[arXiv:2001.01164 [hep-ex]].
	    
	    
\bibitem{Aiola:2020yam}
S.~Aiola, L.~Bandiera, G.~Cavoto, F.~De Benedetti, J.~Fu, V.~Guidi, L.~Henry, D.~Marangotto, F.~M.~Vidal and V.~Mascagna, \textit{et al.}
[arXiv:2010.11902 [hep-ex]]; and references therein.

  \bibitem{Fomin:2019wuw}
		A.~S.~Fomin, S.~Barsuk, A.~Y.~Korchin, V.~A.~Kovalchuk, E.~Kou, M.~Liul, A.~Natochii, E.~Niel, P.~Robbe and A.~Stocchi,
		Eur. Phys. J. C \textbf{80}, no.5, 358 (2020)
		doi:10.1140/epjc/s10052-020-7891-0
		[arXiv:1909.04654 [hep-ph]].


\bibitem{Audurier:2021wqk}
B.~Audurier [LHCb],
Nucl. Phys. A \textbf{1005}, 122001 (2021)
doi:10.1016/j.nuclphysa.2020.122001


\bibitem{Kotulla:2002cg}
M.~Kotulla, J.~Ahrens, J.~R.~M.~Annand, R.~Beck, G.~Caselotti, L.~S.~Fog, D.~Hornidge, S.~Janssen, B.~Krusche and J.~C.~McGeorge, \textit{et al.}
Phys. Rev. Lett. \textbf{89}, 272001 (2002)
doi:10.1103/PhysRevLett.89.272001
[arXiv:nucl-ex/0210040 [nucl-ex]].

\bibitem{Wallace:1995pf}
N.~B.~Wallace, P.~M.~Border, D.~P.~Ciampa, G.~Guglielmo, K.~J.~Heller, D.~M.~Woods, K.~A.~Johns, Y.~T.~Gao, M.~J.~Longo and R.~Rameika,
Phys. Rev. Lett. \textbf{74}, 3732-3735 (1995)
doi:10.1103/PhysRevLett.74.3732

\bibitem{Choudhury:1976dn}
A.~L.~Choudhury and V.~Joshi,
Phys. Rev. D \textbf{13}, 3115-3119 (1976)
doi:10.1103/PhysRevD.13.3115

\bibitem{Choudhury:1976dp}
A.~L.~Choudhury and V.~Joshi,
Phys. Rev. D \textbf{13}, 3120-3124 (1976)
doi:10.1103/PhysRevD.13.3120


\bibitem{Lichtenberg:1976fi}
D.~B.~Lichtenberg,
Phys. Rev. D \textbf{15}, 345 (1977)
doi:10.1103/PhysRevD.15.345

\bibitem{Dattoli:1977nr}
G.~Dattoli, G.~Matone and D.~Prosperi,
Nuovo Cim. A \textbf{45}, 187 (1978)
doi:10.1007/BF02724663

 	\bibitem{Bose:1980vy} 
		S.~K.~Bose and L.~P.~Singh,
		Phys.\ Rev.\ D {\bf 22}, 773 (1980).
		doi:10.1103/PhysRevD.22.773

	\bibitem{Franklin:1981rc}
    J.~Franklin, D.~B.~Lichtenberg, W.~Namgung and D.~Carydas,
    Phys. Rev. D \textbf{24}, 2910 (1981)
    doi:10.1103/PhysRevD.24.2910
	
	\bibitem{Barik:1984tq} 
		N.~Barik and M.~Das,
		Phys.\ Rev.\ D {\bf 28}, 2823 (1983).
		doi:10.1103/PhysRevD.28.2823.
	
		
\bibitem{Morpurgo:1989my}
G.~Morpurgo,
Phys. Rev. D \textbf{40}, 2997 (1989)
doi:10.1103/PhysRevD.40.2997
 
		
		\bibitem{Dey:1994qi}
		J.~Dey, V.~Shevchenko, P.~Volkovitsky and M.~Dey,
		Phys. Lett. B \textbf{337}, 185-188 (1994)
		doi:10.1016/0370-2693(94)91466-4
	
	
		\bibitem{Faessler:2006ft} 
		A.~Faessler, T.~Gutsche, M.~A.~Ivanov, J.~G.~Korner, V.~E.~Lyubovitskij, D.~Nicmorus and K.~Pumsa-ard,
		Phys.\ Rev.\ D {\bf 73}, 094013 (2006)
		doi:10.1103/PhysRevD.73.094013
		[hep-ph/0602193].
		
	
		\bibitem{Albertus:2006ya} 
		C.~Albertus, E.~Hernandez, J.~Nieves and J.~M.~Verde-Velasco,
		Eur.\ Phys.\ J.\ A {\bf 32}, 183 (2007)
		Erratum: [Eur.\ Phys.\ J.\ A {\bf 36}, 119 (2008)]
		doi:10.1140/epja/i2007-10364-y, 10.1140/epja/i2008-10547-0
		[hep-ph/0610030].
		
		
		\bibitem{Roberts:2007ni}
		W.~Roberts and M.~Pervin,
		Int. J. Mod. Phys. A \textbf{23}, 2817-2860 (2008)
		doi:10.1142/S0217751X08041219
		[arXiv:0711.2492 [nucl-th]].
	 
	 
		\bibitem{Sharma:2010vv}
		N.~Sharma, H.~Dahiya, P.~Chatley and M.~Gupta,
		Phys. Rev. D \textbf{81}, 073001 (2010)
		doi:10.1103/PhysRevD.81.073001
		[arXiv:1003.4338 [hep-ph]].

    	
		\bibitem{Albertus:2010hi} 
		C.~Albertus, E.~Hernandez and J.~Nieves,
		Phys.\ Lett.\ B {\bf 690}, 265 (2010)
		doi:10.1016/j.physletb.2010.05.042
		[arXiv:1004.3154 [hep-ph]].


		\bibitem{Bernotas:2012nz} 
		A.~Bernotas and V.~Simonis,
		arXiv:1209.2900 [hep-ph].


		\bibitem{Bernotas:2013eia} 
		A.~Bernotas and V.~Šimonis,
		Phys.\ Rev.\ D {\bf 87}, no. 7, 074016 (2013)
		doi:10.1103/PhysRevD.87.074016
		[arXiv:1302.5918 [hep-ph]].
		
	    
		\bibitem{Simonis:2018rld}
		V.~Simonis,
		arXiv:1803.01809 [hep-ph].


\bibitem{Li:2017cfz}
H.~S.~Li, L.~Meng, Z.~W.~Liu and S.~L.~Zhu,
Phys. Rev. D \textbf{96}, no.7, 076011 (2017)
doi:10.1103/PhysRevD.96.076011
[arXiv:1707.02765 [hep-ph]].


\bibitem{Meng:2017dni}
L.~Meng, H.~S.~Li, Z.~W.~Liu and S.~L.~Zhu,
Eur. Phys. J. C \textbf{77}, no.12, 869 (2017)
doi:10.1140/epjc/s10052-017-5447-8
[arXiv:1710.08283 [hep-ph]].

        \bibitem{Meng:2018gan} 
		L.~Meng, G.~J.~Wang, C.~Z.~Leng, Z.~W.~Liu and S.~L.~Zhu,
		Phys.\ Rev.\ D {\bf 98}, no. 9, 094013 (2018)
		doi:10.1103/PhysRevD.98.094013
		[arXiv:1805.09580 [hep-ph]].
		

		\bibitem{Wang:2018cre} 
		G.~J.~Wang, L.~Meng and S.~L.~Zhu,
		Phys.\ Rev.\ D {\bf 99}, no. 3, 034021 (2019)
		doi:10.1103/PhysRevD.99.034021
		[arXiv:1811.06208 [hep-ph]].

        \bibitem{Yang:2018uoj}
		G.~S.~Yang and H.~C.~Kim,
		Phys. Lett. B \textbf{781}, 601-606 (2018)
		doi:10.1016/j.physletb.2018.04.042
		[arXiv:1802.05416 [hep-ph]].
		
		
		\bibitem{Yang:2019tst}
		G.~S.~Yang and H.~C.~Kim,
		Phys. Lett. B \textbf{801}, 135142 (2020)
		doi:10.1016/j.physletb.2019.135142
		[arXiv:1909.03156 [hep-ph]].

        \bibitem{Yang:2020klp}
		G.~S.~Yang and H.~C.~Kim,
		Phys. Lett. B \textbf{808}, 135619 (2020)
		doi:10.1016/j.physletb.2020.135619
		[arXiv:2004.08524 [hep-ph]].
		

		\bibitem{Aliev:2001ig}
		T.~Aliev, A.~Ozpineci and M.~Savci,
		Phys. Rev. D \textbf{65}, 056008 (2002)
		doi:10.1103/PhysRevD.65.056008
		[arXiv:hep-ph/0107196 [hep-ph]];
		
		
		\bibitem{Aliev:2001xr}
		T.~Aliev, A.~Ozpineci and M.~Savci,
		Phys. Rev. D \textbf{65}, 096004 (2002)
		doi:10.1103/PhysRevD.65.096004
		[arXiv:hep-ph/0110074 [hep-ph]].
		
	
		\bibitem{Aliev:2008ay}
		T.~Aliev, K.~Azizi and A.~Ozpineci,
		Phys. Rev. D \textbf{77}, 114006 (2008)
		doi:10.1103/PhysRevD.77.114006
		[arXiv:0803.4420 [hep-ph]];
		
	
		\bibitem{Aliev:2008sk} 
		T.~M.~Aliev, K.~Azizi and A.~Ozpineci,
		Nucl.\ Phys.\ B {\bf 808}, 137 (2009)
		doi:10.1016/j.nuclphysb.2008.09.018
		[arXiv:0807.3481 [hep-ph]];
		
		\bibitem{Aliev:2009jt} 
		T.~M.~Aliev, K.~Azizi and A.~Ozpineci,
		Phys.\ Rev.\ D {\bf 79}, 056005 (2009)
		doi:10.1103/PhysRevD.79.056005
		[arXiv:0901.0076 [hep-ph]].
		
		
		\bibitem{Aliev:2014bma} 
		T.~M.~Aliev, K.~Azizi and H.~Sundu,
		Eur.\ Phys.\ J.\ C {\bf 75}, no. 1, 14 (2015)
		doi:10.1140/epjc/s10052-014-3229-0
		[arXiv:1409.7577 [hep-ph]].
		
		
		\bibitem{Aliev:2016xvq} 
		T.~M.~Aliev, T.~Barakat and M.~Savcı,
		Phys.\ Rev.\ D {\bf 93}, no. 5, 056007 (2016)
		doi:10.1103/PhysRevD.93.056007
		[arXiv:1603.04762 [hep-ph]].
		
		
		\bibitem{Liu:2009jc}
		L.~Liu, H.~W.~Lin, K.~Orginos and A.~Walker-Loud,
		Phys. Rev. D \textbf{81}, 094505 (2010)
		doi:10.1103/PhysRevD.81.094505
		[arXiv:0909.3294 [hep-lat]].

	  	
		\bibitem{Brown:2014ena} 
		Z.~S.~Brown, W.~Detmold, S.~Meinel and K.~Orginos,
		Phys.\ Rev.\ D {\bf 90}, no. 9, 094507 (2014)
		doi:10.1103/PhysRevD.90.094507
		[arXiv:1409.0497 [hep-lat]].

        \bibitem{Can:2013tna} 
		K.~U.~Can, G.~Erkol, B.~Isildak, M.~Oka and T.~T.~Takahashi,
		JHEP {\bf 1405}, 125 (2014)
		doi:10.1007/JHEP05(2014)125
		[arXiv:1310.5915 [hep-lat]].

	   	\bibitem{Bahtiyar:2018vub}
		H.~Bahtiyar, K.~U.~Can, G.~Erkol, M.~Oka and T.~T.~Takahashi,
		Phys. Rev. D \textbf{98}, no.11, 114505 (2018)
		doi:10.1103/PhysRevD.98.114505
		[arXiv:1807.06795 [hep-lat]].
       
        
		\bibitem{Mathur:2018rwu}
		N.~Mathur and M.~Padmanath,
		Phys. Rev. D \textbf{99}, no.3, 031501 (2019)
		doi:10.1103/PhysRevD.99.031501
		[arXiv:1807.00174 [hep-lat]].

	\bibitem{Karliner:2014gca}
    M.~Karliner and J.~L.~Rosner,
Phys. Rev. D \textbf{90}, no.9, 094007 (2014)
doi:10.1103/PhysRevD.90.094007
[arXiv:1408.5877 [hep-ph]].
	

\bibitem{Karliner:2017gml}
M.~Karliner and J.~L.~Rosner,
Phys. Rev. D \textbf{96}, no.3, 033004 (2017)
doi:10.1103/PhysRevD.96.033004
[arXiv:1706.06961 [hep-ph]].
		
		\bibitem{Karliner:2018hos}
		M.~Karliner and J.~L.~Rosner,
		Phys. Rev. D \textbf{97}, no.9, 094006 (2018)
		doi:10.1103/PhysRevD.97.094006
		[arXiv:1803.01657 [hep-ph]].

	  	\bibitem{Wang:2017kfr}
		K.~L.~Wang, Y.~X.~Yao, X.~H.~Zhong and Q.~Zhao,
		Phys. Rev. D \textbf{96}, no.11, 116016 (2017)
		doi:10.1103/PhysRevD.96.116016
		[arXiv:1709.04268 [hep-ph]].
		
	
		\bibitem{Majethiya:2009vx} 
		A.~Majethiya, B.~Patel and P.~C.~Vinodkumar,
		Eur.\ Phys.\ J.\ A {\bf 42}, 213 (2009)
		doi:10.1140/epja/i2009-10880-8
		[arXiv:0902.2536 [hep-ph]].
	    
	   
		\bibitem{Shah:2016vmd}
		Z.~Shah, K.~Thakkar and A.~K.~Rai,
		Eur. Phys. J. C \textbf{76}, no.10, 530 (2016)
		doi:10.1140/epjc/s10052-016-4379-z
		[arXiv:1609.03030 [hep-ph]].

		\bibitem{Gandhi:2018lez}
		K.~Gandhi, Z.~Shah and A.~K.~Rai,
		Eur. Phys. J. Plus \textbf{133}, no.12, 512 (2018)
		doi:10.1140/epjp/i2018-12318-1
		[arXiv:1811.00251 [hep-ph]].
		
		
		
		\bibitem{Li:2017pxa}
		H.~S.~Li, L.~Meng, Z.~W.~Liu and S.~L.~Zhu,
		Phys. Lett. B \textbf{777}, 169-176 (2018)
		doi:10.1016/j.physletb.2017.12.031
		[arXiv:1708.03620 [hep-ph]].
		
		
		\bibitem{Lu:2017meb}
		Q.~F.~L\"u, K.~L.~Wang, L.~Y.~Xiao and X.~H.~Zhong,
		Phys. Rev. D \textbf{96}, no.11, 114006 (2017)
		doi:10.1103/PhysRevD.96.114006
		[arXiv:1708.04468 [hep-ph]].
		
		\bibitem{Xiabng:2018qsd} 
		R.~X.~Shi, Y.~Xiao and L.~S.~Geng,
		Phys.\ Rev.\ D {\bf 100}, no. 5, 054019 (2019)
		doi:10.1103/PhysRevD.100.054019
		[arXiv:1812.07833 [hep-ph]].
		
		\bibitem{Liu:2018euh} 
		M.~Z.~Liu, Y.~Xiao and L.~S.~Geng,
		Phys.\ Rev.\ D {\bf 98}, no. 1, 014040 (2018)
		doi:10.1103/PhysRevD.98.014040
		[arXiv:1807.00912 [hep-ph]].
		
		\bibitem{Ozdem:2018uue} 
		U.~Özdem,
		J.\ Phys.\ G {\bf 46}, no. 3, 035003 (2019)
		doi:10.1088/1361-6471/aafffc
		[arXiv:1804.10921 [hep-ph]].
		
		\bibitem{Ozdem:2019zis}
		U.~\"Ozdem,
		Eur. Phys. J. A \textbf{56}, no.2, 34 (2020)
		doi:10.1140/epja/s10050-020-00049-4
		[arXiv:1906.08353 [hep-ph]].
		
		\bibitem{Kumar:2005ei}
		S.~Kumar, R.~Dhir and R.~C.~Verma,
		J. Phys. G \textbf{31}, no.2, 141-147 (2005)
		doi:10.1088/0954-3899/31/2/006
		
		
		\bibitem{Dhir:2009ax}
		R.~Dhir and R.~Verma,
		Eur. Phys. J. A \textbf{42}, 243-249 (2009)
		doi:10.1140/epja/i2009-10872-8
		[arXiv:0904.2124 [hep-ph]].
		
		\bibitem{Dhir:2013nka}
		R.~Dhir, C.~S.~Kim and R.~C.~Verma,
		Phys. Rev. D \textbf{88}, 094002 (2013)
		doi:10.1103/PhysRevD.88.094002
		[arXiv:1309.4057 [hep-ph]].
	
		
\bibitem{Dillon:1995qw}
G.~Dillon and G.~Morpurgo,
Phys. Rev. D \textbf{53}, 3754-3769 (1996)
doi:10.1103/PhysRevD.53.3754


\bibitem{Durand:2001zz}
L.~Durand, P.~Ha and G.~Jaczko,
Phys. Rev. D \textbf{64}, 014008 (2001)
doi:10.1103/PhysRevD.64.014008
[arXiv:hep-ph/0101267 [hep-ph]].


\bibitem{Durand:2001sz}
L.~Durand, P.~Ha and G.~Jaczko,
Phys. Rev. D \textbf{65}, 034019 (2002)
[erratum: Phys. Rev. D \textbf{65}, 099904 (2002)]
doi:10.1103/PhysRevD.65.099904
[arXiv:hep-ph/0104197 [hep-ph]].


\end{thebibliography}
\end{document}